\documentclass{mnras}

\usepackage{graphicx}
\usepackage{footnote}
\usepackage[font=small]{caption}
\usepackage{indentfirst}
\usepackage{tabu}
\usepackage[percent]{overpic}
\usepackage{transparent}
\usepackage{ragged2e}
\usepackage[utf8]{inputenc}
\usepackage{multirow}
\usepackage{cmap}
\usepackage[T1]{fontenc}
\usepackage[normalem]{ulem} 

\let\bibhang\relax
\usepackage{natbib}
\usepackage{booktabs}
\usepackage{fixltx2e}

\usepackage{url}
\usepackage{hyperref}

\usepackage{amssymb}
\usepackage{txfonts}
\usepackage[a]{esvect}
\usepackage{tikz}

\newcommand{\degree}[0]{$^{\circ}$}

\let\baraccent=\=
\renewcommand{\=}[1]{\stackrel{#1}{=}} 
\let\arrowaccent=\>
\renewcommand{\>}[1]{\stackrel{#1}{\Rightarrow}} 

\makeatletter
\newcommand{\rmnum}[1]{{\footnotesize{\expandafter\@slowromancap\romannumeral #1@}}}
\newcommand{\Rmnum}[1]{{\expandafter\@slowromancap\romannumeral #1@}}
\makeatother

\newcommand{\tfm}[1]{$^{#1}$}
\newcommand{\tablefoottext}[2]{$^{(#1)}$ #2}

\hypersetup{
   colorlinks=true,       
   linkcolor=black,       
   citecolor=blue,        
   urlcolor=black         
}

\begin{document}

\title[SFE and AGN feedback in NLSy1]{Star formation efficiency and AGN feedback in narrow-line Seyfert 1 galaxies with fast X-ray nuclear winds}

\author[Salom\'e et al.]{
   Q. Salom\'e$^{1,2}$,
   Y. Krongold$^{3}$,
   A. L. Longinotti$^{3}$,
   M. Bischetti$^{4,5}$,
   S. Garc\'ia-Burillo$^{6}$,
   \newauthor
   O. Vega$^{7}$,
   M. S\'anchez-Portal$^{8}$,
   C. Feruglio$^{5,9}$,
   M. J. Jim\'enez-Donaire$^{6,10}$,
   M. V. Zanchettin$^{11,5}$
\\
   $^{1}$  Finnish Centre for Astronomy with ESO (FINCA), University of Turku, Vesilinnantie 5, 20014 Turku, Finland \\
   email: quentin.salome@utu.fi \\
   $^{2}$  Aalto University Mets\"ahovi Radio Observatory, Mets\"ahovintie 114, 02540 Kylm\"al\"a, Finland \\
   $^{3}$  Instituto de Astronom\'ia, Universidad Nacional Aut\'onoma de M\'exico, AP 70-264, Ciudad de M\'exico, CDMX 04510, Mexico \\
   $^{4}$  Dipartimento di Fisica, Universit\'a di Trieste, Sezione di Astronomia, Via G. Tiepolo 11, 34131 Trieste, Italy \\
   $^{5}$  INAF Osservatorio Astronomico di Trieste, via G. Tiepolo 11, 34143 Trieste, Italy \\
   $^{6}$  Observatorio Astron\'omico Nacional (IGN), C/Alfonso XII 3, 28014 Madrid, Spain \\
   $^{7}$  Instituto Nacional de Astrof\'isica, \'Optica y Electr\'onica, Luis E. Erro 1, Tonantzintla 72840, Puebla , Mexico \\
   $^{8}$  Institut de Radioastronomie Millim\'etrique, Avenida Divina Pastora 7, Local 20, 18012 Granada, Spain \\
   $^{9}$  IFPU-Institute for Fundamental Physics of the Universe, via Beirut 2, 34014 Trieste, Italy \\
   $^{10}$ Centro de Desarrollos Tecnol\'ogicos, Observatorio de Yebes (IGN), 19141 Yebes, Guadalajara, Spain \\
   $^{11}$ SISSA, Via Bonomea 265, 34136 Trieste, Italy
}

\date{Accepted 2023 June 25. Received 2023 June 22; in original form 2022 July 12}

\pubyear{2020}

\label{firstpage}
\pagerange{\pageref{firstpage}--\pageref{lastpage}}
\maketitle

\begin{abstract}
   We present the first systematic study of the molecular gas and star formation efficiency in a sample of ten narrow-line Seyfert 1 galaxies selected to have X-ray Ultra Fast Outflows and, therefore, to potentially show AGN feedback effects. CO observations were obtained with the IRAM 30m telescope in six galaxies and from the literature for four galaxies. We derived the stellar mass, star formation rate, AGN and FIR dust luminosities by fitting the multi-band spectral energy distributions with the CIGALE code. Most of the galaxies in our sample lie above the main sequence (MS) and the molecular depletion time is one to two orders of magnitude shorter than the one typically measured in local star-forming galaxies. Moreover, we found a promising correlation between the star formation efficiency and the Eddington ratio, as well as a tentative correlation with the AGN luminosity. The role played by the AGN activity in the regulation of star formation within the host galaxies of our sample remains uncertain (little or no effect? positive feedback?). Nevertheless, we can conclude that quenching by the AGN activity is minor and that star formation will likely stop in a short time due to gas exhaustion by the current starburst episode.
\end{abstract}



\begin{table*}
  \centering
  \caption{\label{table:sample} Sample of NLSy1 galaxies with known X-ray UFO. The columns correspond to: (1) Name of the galaxy; (2-3) central coordinates; (4) spectroscopic redshift; (5) luminosity distance; (6) AGN bolometric luminosity estimated by fitting the multi-band SED with the code CIGALE (see Section \ref{sec:cigale} and Table \ref{table:outcigale}); (7) mass of the central black hole; (8) Eddigton ratio $\lambda_{Edd}=L_{AGN}/L_{Edd}$ where $L_{Edd}$ is the Eddington luminosity; (10) velocity of the X-ray UFO, and (11) references for the black hole mass and UFO velocity. The demarcation line separates the sources observed with the IRAM 30m from those with ancillary CO data.}
  \begin{tabular}{lcrccccccl}
    \hline \hline
    Galaxy           &      RA      & \multicolumn{1}{c}{Dec} &       z        & $D_L$ &  $log L_{AGN}$  & $log M_{BH}$ & $\lambda_{Edd}$ &  $v_{UFO}$   & Ref.   \\
                     &              &                         &                & (Mpc) & ($erg\,s^{-1}$) & ($M_\odot$)  &                 &              &        \\ \hline
    IRAS\,13349+2438 & 13:37:18.727 &       24:23:03.38       & 0.10852\tfm{a} & 502.4 &      46.00      &     8.63     &      0.19       &    0.14c     & 1,2    \\
    IRAS\,17020+4544 & 17:03:30.383 &       45:40:47.17       & 0.0612\tfm{a}  & 274.3 &      44.93      &     6.77     &      1.14       &    0.11c     & 3      \\
    Mrk\,205         & 12:21:44.221 &       75:18:38.83       &    0.07085     & 319.7 &      44.87      &     8.32     &      0.03       &    0.14c     & 2,4    \\
    Ark\,564         & 22:42:39.345 &       29:43:31.31       &    0.02468     & 107.7 &      44.34      &     6.27     &      0.93       &    0.16c     & 5,6,7  \\
    Mrk\,1044        & 02:30:05.525 &    $-$08:59:53.29       &    0.01645     &  71.3 &      44.06      &     6.45     &      0.32       &    0.15c     & 8      \\
    Mrk\,110         & 09:25:12.870 &       52:17:10.52       &    0.03529     & 155.2 &      44.15      &     7.29     &      0.06       & 0.13c\tfm{b} & 7      \\ \hline
    I\,Zw\,1         & 00:53:34.940 &       12:41:36.20       &    0.06114     & 274.0 &      45.43      &     7.46     &      0.74       &    0.21c     & 2,9,10 \\
    PG\,1211+143     & 12:14:17.670 &       14:03:13.10       &    0.08090     & 367.6 &      45.62      &     7.61     &      0.80       &    0.13c     & 4,7    \\
    PG\,1448+273     & 14:51:08.763 &       27:09:26.92       &    0.06449     & 289.7 &      44.95      &     7.29     &      0.36       &    0.15c     & 11     \\
    Mrk\,766         & 12:18:26.509 &       29:48:46.34       &    0.01288     &  55.7 &      43.83      &     6.82     &      0.08       &    0.08c     & 4,7    \\ \hline
  \end{tabular} \\
  \justify {\small
  \textbf{Notes.} 
    \tablefoottext{a}{CO-estimated redshift (see \citealt{SalomeQ_2021} and Section \ref{sec:CO}).}
    \tablefoottext{b}{Segura-Montero et al. (in prep.)} \\
  \textbf{References.} (1) \cite{Parker_2018}, (2) \cite{Afanasiev_2019}, (3) \cite{Longinotti_2015}, (4) \cite{Tombesi_2010}, (5) \cite{Gupta_2013}, (6) \cite{Igo_2020}, (7) \cite{Waddell_2020}, (8) \cite{Krongold_2021}, (9) \cite{Cicone_2014}, (9) \cite{Reeves_2019}, (11) \cite{Laurenti_2021}.
  }
\end{table*}


\section{Introduction}
\label{sec:introduction}

   It is now well-established that active galactic nuclei (AGN) and their hosts galaxies are evolving together, influencing each other (see the review by \citealt{Heckman_2014}). In particular, AGN feedback is often invoked as a key element for star formation and galaxy evolution \citep{diMatteo_2005,Hopkins_2010}. However, the mechanisms governing AGN feedback are still not understood. Galaxy outflows and winds driven by the AGN have the potential to impact star formation.
Several models (e.g. \citealt{Faucher-Giguere_2012,King_2015}) have proposed that quasar feedback is initiated via a sub-relativistic wind launched at the scale of the accretion disc with velocity $v>10^4\: \rm km\,s^{-1}$ and observed in X-ray spectra as ultra-fast outflows (UFO; \citealt{Tombesi_2010,Tombesi_2012}).

   While traveling outward, the ultra-fast wind interacts with the surrounding interstellar medium (ISM), producing large scale outflows that are commonly observed in the optical band \citep{Marasco_2020,Robleto_2021} and at millimetric frequencies \citep{Feruglio_2010,Feruglio_2015,Cicone_2012,Cicone_2014,Tombesi_2015,Bischetti_2019, Zanchettin_2021,Longinotti_2023}. At the contact discontinuity produced in the shock of the wind with the ISM, the pressure and momentum of the wind are conserved. The interaction also produces an inner reverse shock front where the nuclear wind slows, and an outer forward shock which accelerates the ISM (see the review by \citealt{King_2015}).
If the inner and outer shock fronts are associated with cooling, a large fraction of the kinetic energy of the nuclear wind is dissipated (momentum-conserving outflow). Conversely, if the shocked regions are expanding semi-adiabatically, most of the kinetic energy is transmitted to the large-scale outflow (energy-conserving outflow) and the momentum flux at large scale is boosted \citep{King_2015}.

   A common method to study the coupling of the nuclear wind with the galaxy-scale outflow is to look at the energetics of the X-ray and of the molecular phases (e.g. \citealt{Feruglio_2015,Longinotti_2023}). Few sources show the presence of both an X-ray UFO and a molecular outflow (e.g. Mrk\,509; \citealt{Tombesi_2010,Zanchettin_2021}). In four cases, the momentum flux of both phases shows that the outflows seem to be driven by an energy-conserving wind, while for the others sources they are consistent with the momentum-conserving scenario. \cite{Marasco_2020} suggested that these two well-defined regimes trace an evolutionary path from highly accreting sources, which are still growing their black holes, to a phase where the outflows can finally eject most of the ISM by reaching galaxy scales, quenching both star formation processes and black hole growth. 

   In this regard, narrow-line Seyfert 1 galaxies (NLSy1) represent an interesting class of sources where to explore feedback properties. NLSy1 are young AGN characterised by a high accretion rate (Eddington ratio larger than 0.1; \citealt{Boroson_1992,Mathur_2000a,Mathur_2000b}) on supermassive black holes with generally small masses ($M_{BH}<10^{8} M_{\odot}$; \citealt{Peterson_2011}).
Recent findings of X-ray fast outflowing gas \citep{Gupta_2013,Longinotti_2015,Parker_2017,Krongold_2021} and relativistic jets \citep{Lahteenmaki_2017, Lahteenmaki_2018} suggest that nuclear winds may be a common feature of this AGN class possibly due to their high accretion rate.
Although the bolometric luminosity of NLSy1 is rather moderate compared to the typical dust and gas-rich sources where feedback processes are expected (ULIRGs and quasars; \citealt{Cicone_2014,Jarvela_2015}), the presence of X-ray UFO may play a role in triggering galaxy scale outflows capable of producing efficient feedback.

   In this paper, we seek for any possible effect indicating that the presence of nuclear fast outflow winds may imprint in the properties of the host galaxy at large scale. We therefore characterise the molecular gas content (as traced by the CO emission) and study the star formation efficiency (as traced by the molecular depletion time) in a sample of NLSy1 for which a X-ray UFO was clearly observed. We aim to establish whether the molecular gas is affected or not by the presence of the AGN-driven winds. The sample and the observations are presented in Section \ref{sec:Obs} and analysed in Section \ref{sec:Res}. We then discuss the relation between the molecular gas reservoir, the stellar mass and the star formation in Section \ref{sec:gas-SF}, as well as the possible impact of the AGN activity in Section \ref{sec:AGN-feedback}. Finally, Section \ref{sec:conclusion} will summarise the results.
Throughout this paper, we assume $H_0=70$, $\Omega_M=0.3$, $\Omega_{vac}=0.7$.

\section{Observations}
\label{sec:Obs}

   \subsection{Sample}

   We compiled our sample with all the known NLSy1 for which nuclear X-ray fast winds are well established. We also limit our sample to the sources observable from the IRAM 30m telescope (declination above -15 degrees). We selected the following sources: PG\,1211+143, Mrk\,205, Mrk\,766 from positive detections in the sample of \cite{Tombesi_2010}; Ark\,564, I\,Zw\,1, PG\,1448+273, IRAS\,13349+2438, Mrk\,1044, IRAS\,17020+4544 (hereafter IRAS\,17020) all from the literature on individual objects; and Mrk\,110 from our ongoing work on X-ray winds. The selected sample consists of 10 sources. The references for the UFO in each of them are given in Table \ref{table:sample}.

  Five of these sources were previously observed in CO: IRAS\,17020 for which the molecular gas at systemic velocity and in outflows was studied with NOEMA by \cite{SalomeQ_2021} and \cite{Longinotti_2023}; I\,Zw\,1 which is barely resolved in CO(1-0) with NOEMA \citep{Cicone_2014}; Mrk\,766 in which most of the CO(2-1) emission is concentrated in the inner 0.5 kpc \citep{Dominguez-Fernandez_2020}; PG\,1211+143 and PG\,1448+273 (hereafter PG\,1211 and PG\,1448) which were observed in CO(2-1) with ACA \citep{Shangguan_2020a}.

   With the IRAM 30m, we observed the five northern sources with no previous CO observations: IRAS\,13349+2438 (hereafter IRAS\,13449), Ark\,564, Mrk\,110, Mrk\,205 and Mrk\,1044. We also re-observed IRAS\,17020 as a reference target to see whether molecular outflows can be detected or not. However, the data are not sensitive enough to detect the broad emission observed with the LMT and NOEMA by \cite{Longinotti_2018,Longinotti_2023}.

   Global properties of the galaxies in our sample are presented in Table \ref{table:sample}. Although limited in size, our sample spans two orders of magnitude in bolometric luminosity and in black hole mass.


\begin{table*}
  \centering
  \caption{\label{table:CO-spec} Properties of the CO emission as observed with the IRAM 30m observations: (1) Short name; (2) on-source observing time; (3) CO transition; (4) central redshifted frequency; (5) noise rms; (6) channel width for which the rms was derived; (7) main beam peak temperature; (8) full-width at half maximum; (9) peak velocity; (10) integrated intensity.
  If the line emission presents a double-peak profile, we characterised each peak with a Gaussian and give the values for each peak. We have considered a systematic uncertainty of 10\% on the absolute flux calibration.}
  \begin{tabular}{lclccccccc}
    \hline \hline
    Galaxy      & $t_{obs}$ &  line   & $\nu_{obs}$ &  rms &   $\delta v$   & $T_{mb}$ &      FWHM      &       $v_0$       &      $I_{CO}$     \\
                &   (min)   &         &    (GHz)    & (mK) & ($km\,s^{-1}$) &   (mK)   & ($km\,s^{-1}$) &  ($km\,s^{-1}$)   & ($K\,km\,s^{-1}$) \\ \hline
    IRAS\,13349 &    115    & CO(1-0) &  104.06919  &  0.9 &       20       &   3.22   &   $226\pm 26$  &   $58.7\pm 13.9$  &   $0.77\pm 0.12$  \\ 
                &           &         &             &      &                &   3.18   &   $188\pm 28$  &  $425.6\pm 12.5$  &   $0.64\pm 0.10$  \\ 
                &           & CO(2-1) &  208.13440  &  1.5 &       30       &   5.21   &   $517\pm 51$  &  $198.4\pm 25.6$  &   $2.87\pm 0.40$  \\ 
                \hline                                                        
    IRAS\,17020 &    109    & CO(1-0) &  108.62344  &  1.0 &       20       &   4.56   &   $161\pm 20$  & $-157.9\pm 8.4$   &   $0.78\pm 0.11$  \\ 
                &           &         &             &      &                &   4.87   &   $143\pm 17$  &  $140.8\pm 7.2$   &   $0.74\pm 0.11$  \\ 
                &           & CO(2-1) &  217.24273  &  1.7 &       20       &   10.3   &   $196\pm 24$  & $-160.0\pm 9.9$   &   $2.15\pm 0.31$  \\ 
                &           &         &             &      &                &   10.4   &   $211\pm 25$  &  $142.4\pm 10.1$  &   $2.33\pm 0.33$  \\ 
                \hline                                                        
    Mrk\,205    &    112    & CO(1-0) &  107.64458  &  1.5 &       20       &   7.13   &   $125\pm 14$  &    $9.5\pm 7.0$   &   $0.95\pm 0.14$  \\ 
                \hline                                                        
    Ark\,564    &     61    & CO(1-0) &  112.49482  &  1.9 &       30       &   7.18   &   $163\pm 34$  &   $54.6\pm 13.7$  &   $1.25\pm 0.25$  \\ 
                &           & CO(2-1) &  224.98535  &  3.6 &       25       &   8.70   &   $125\pm 23$  &   $14.6\pm 16.4$  &   $1.20\pm 0.30$  \\ 
                \hline                                                        
    Mrk\,1044   &     92    & CO(1-0) &  113.40567  &  1.7 &       20       &   10.6   &    $65\pm 19$  & $-103.3\pm 7.0$   &   $0.73\pm 0.34$  \\ 
                &           &         &             &      &                &   9.12   &   $147\pm 24$  &  $-13.2\pm 20.1$  &   $1.43\pm 0.38$  \\ 
                &           & CO(2-1) &  226.80701  &  2.7 &       30       &   11.4   &    $58\pm 15$  &  $-85.5\pm 7.9$   &   $0.71\pm 0.17$  \\ 
                &           &         &             &      &                &   8.96   &    $44\pm 29$  &   $42.1\pm 7.8$   &   $0.42\pm 0.16$  \\ 
                \hline                                                        
    Mrk\,110    &    120    & CO(1-0) &  111.34194  &  1.2 &       25       &   5.70   &    $78\pm 13$  &   $13.8\pm 6.6$   &   $0.48\pm 0.09$  \\ 
                &           & CO(2-1) &  222.67962  &  3.0 &       25       &  $<9.0$  &       -        &         -         &      $<0.76$      \\
                \hline
  \end{tabular} \\
  \justify {\small
  \textbf{Notes.} 
    For the upper limit of the CO(2-1) emission in Mrk\,110, we assumed a full-width at half maximum $FWHM=80\: km\,s^{-1}$.
  }
\end{table*}


   \subsection{Observational setup}

   Six NLSy1 galaxies of our sample were observed in CO(1-0) and CO(2-1) simultaneously with the IRAM 30m telescope, with the exception of Mrk\,205 for which the redshift did not allow us to find a configuration to observe both lines simultaneously. Observations were made in May-June 2020 using the EMIR receiver\footnote{\url{http://www.iram.es/IRAMES/mainWiki/EmirforAstronomers}} with the FTS (bandwidth of $2\times 4$ GHz; resolution of 195 kHz) and WILMA backends (bandwidth of 4 GHz; resolution of 2 MHz).
At the redshift of the galaxies, the lines are observable at frequencies of 104-113.5 GHz and 208-227 GHz, which leads to a beam of about $22''-24''$ and $11''-12''$, respectively. The beams of the IRAM 30m thus translate into spatial scales for our sources that go from $3.7-7.4\: kpc$ in Mrk\,1044 to $24-48\: \rm kpc$ in IRAS\,13349. Along the observing nights, the system temperature varied between 100-300 K at 3mm and 200-700 K at 1mm.
During observations, the pointing was monitored by observing standard continuum sources tuned to the frequency corresponding to the redshifted CO(1-0) emission line. Observations were obtained using wobbler switching with a rate of $\sim 0.5\: \rm Hz$. Six-minute scans were taken, and a calibration was made every three scans. The observing time on-source varies between 60-120 min, providing a noise level of $\sim 0.9-1.7\: \rm mK$ in CO(1-0) and $\sim 1.5-3.7\: \rm mK$ in CO(2-1) in channels of $20-30\: \rm km\,s^{-1}$ (see Table \ref{table:CO-spec} for the details). Pointing was checked every few hours and was generally determined to be accurate to within a few arcseconds.

\section{Results}
\label{sec:Res}

   \subsection{CO emission from IRAM 30m}
   \label{sec:CO}

   The data were reduced and analysed using the IRAM package {\tt CLASS}\footnote{\url{https://www.iram.fr/IRAMFR/GILDAS/}}. The average spectra were first smoothed to a spectral resolution of $20-30\: \rm km\,s^{-1}$ (refer to Table \ref{table:CO-spec} for the details). The baseline was then subtracted by fitting the channels with no signal with a linear function or a degree 2 polynomial. In the case of the CO(1-0) spectrum in Mrk\,1044, a platform is observed with a turnover velocity $\sim 320\: \rm km\,s^{-1}$, outside the velocity range of the CO emission. This was corrected by fitting a baseline for each platform separately. The resulting spectra are plotted in Appendix \ref{sec:spectra}.
The CO(1-0) emission is detected in all six galaxies with a signal-to-noise ratio $\geq 3.5$ for the peak temperature. Regarding the CO(2-1) line, it was not detected in Mrk\,110. The CO emission presents a double-horn profile in IRAS\,13349, IRAS\,17020 and Mrk\,1044, suggesting the presence of molecular gas disc. On the other hand, only one gaussian-like profile is observed in the other galaxies. We fitted the CO emission using one or two Gaussians. The characteristics of the emission lines are summarised in Table \ref{table:CO-spec}.
\medskip

\noindent \textit{IRAS\,17020+4544} - The CO(1-0) emission was previously observed with the LMT \citep{Longinotti_2018} and NOEMA \citep{SalomeQ_2021}. With the IRAM 30m, we found a CO luminosity in the host galaxy, in agreement with both the LMT and NOEMA measurements. We thus confirm the conclusion of \cite{SalomeQ_2021} that all the CO emission was recovered by the NOEMA observations, with no spatial filtering by the interferometer. However, the data from the IRAM 30m show no evidence for the molecular outflows detected with the LMT \citep{Longinotti_2018} and NOEMA \citep{Longinotti_2023}. The rms noise of 3.9 mJy we reached with the IRAM 30m for CO(1-0) is higher than the peak temperature of $\lesssim 1\: \rm mJy$ for the resolved outflows with NOEMA and 1.1 mJy for the broad component observed with the LMT. Therefore, the non detection of the outflow components is simply due to a lack of sensitivity.
\medskip

\noindent \textit{IRAS\,13349+2438} - The CO(1-0) spectrum presents a double-horn profile, while the CO(2-1) emission does not. A plausible explanation is that it is a result of the difference in the beam size. In particular, the beam of the IRAM 30m for the CO(2-1) is about half the optical diameter of the galaxy and the CO(2-1) emission may not be fully encompassed with one beam. Therefore, a significant fraction of the CO(2-1) can potentially be missed by the IRAM 30m. However, we cannot exclude that the gas close to the centre is more excited and presents brighter CO(2-1) emission than the outskirt region.

   We note that the two peaks in the CO(1-0) spectrum of IRAS\,13349 are redshifted compared with the optically estimated redshift ($z_{opt}=0.107641$ based on the $H\alpha$, $H\beta$ and [O\rmnum{3}] lines; \citealt{Kim_1995}). Using the profile of the CO(1-0) emission, we find a new systemic velocity (corresponding to the centre of the double-peak) redshifted by $240\: \rm km\,s^{-1}$ from the nominal frequency. This corresponds to a redshifted frequency of 103.98596 GHz. We thus get a CO-estimated redshift $z_{CO}=0.10852$, which corresponds to a luminosity distance of 502.4 Mpc.
Differences between the systemic velocity of the CO lines and optical lines are commonly observed.
Such difference may be produced by outflowing gas within the narrow-line region or dynamical perturbations due to galaxy interactions (e.g. \citealt{Davies_2020b}).
We note that the velocity difference between the optical and CO emission in IRAS\,13349 is similar to the velocity difference observed in IRAS\,17020 \citep{SalomeQ_2021}. This suggests that the same process may be at play in both galaxies.

   Interferometric observations of the molecular gas and integral field spectroscopy of IRAS\,13349 will be necessary to investigate both the different profiles of the CO(1-0) and CO(2-1) emission, and the velocity difference between the CO and optical emission.


\begin{table}
  \centering
  \caption{\label{table:LCO-MH2} Derived properties of the CO emission: (1) CO luminosity corrected for aperture effect; (2) line ratio in the hypothesis of unresolved emission; (3) molecular gas mass derived from the CO luminosity with the CO-to-H$_2$ conversion factor in Section \ref{sec:H2_mass}. When the spectrum presents a double-horn profile, the CO luminosity is the sum of the two Gaussians. The demarcation line separates the sources observed with the IRAM 30m from those with ancillary CO data.}
  \begin{tabular}{lccc}
    \hline \hline
    Galaxy      &       $L'_{CO10}$         &    $R_{CO21}$    &         $M_{H_2}$         \\
                &  ($K\,km\,s^{-1}\,pc^2$)  &                  &        ($M_\odot$)        \\ \hline
    IRAS\,13349 & $(4.1\pm 0.7)\times 10^9$ & $(0.51\pm 0.15)$ & $(3.3\pm 0.6)\times 10^9$ \\
    IRAS\,17020 & $(1.5\pm 0.3)\times 10^9$ & $(0.74\pm 0.21)$ & $(1.2\pm 0.2)\times 10^9$ \\
    Mrk\,205    & $(1.0\pm 0.1)\times 10^9$ &        -         & $(4.3\pm 0.4)\times 10^9$ \\
    Ark\,564    & $(2.0\pm 0.4)\times 10^8$ & $(0.24\pm 0.11)$ & $(1.6\pm 0.3)\times 10^8$ \\
    Mrk\,1044   & $(1.5\pm 0.5)\times 10^8$ & $(0.13\pm 0.08)$ & $(6,5\pm 2.2)\times 10^8$ \\
    Mrk\,110    & $(1.8\pm 0.3)\times 10^8$ &     $<0.40$      & $(7.7\pm 1.3)\times 10^8$ \\ \hline
    I\,Zw\,1    &  $5.5\times 10^9$\tfm{a}  &   $0.63$\tfm{b}  &     $4.4\times 10^9$      \\
    PG\,1211    &  $9.3\times 10^7$\tfm{b}  &  $>0.11$\tfm{c}  &     $7.4\times 10^7$      \\
    PG\,1448    &  $3.8\times 10^8$\tfm{b}  &        -         &     $1.6\times 10^9$      \\
    Mrk\,766    &  $3.8\times 10^8$\tfm{d}  &        -         &     $3.0\times 10^8$      \\ \hline
  \end{tabular} \\
  \justify {\small
  \textbf{Notes.} 
    \tablefoottext{a}{\cite{Evans_2006,Cicone_2014}}
    \tablefoottext{b}{\cite{Shangguan_2020a}}
    \tablefoottext{c}{Line ratio $L'_{CO21}/L'_{CO10}$ from \cite{Shangguan_2020a}}
    \tablefoottext{d}{Derived from \cite{Dominguez-Fernandez_2020}}
  }
\end{table}


   \subsection{CO line ratio}
   \label{sec:line_ratio}

   In this section, we look at the integrated intensities line ratio when both CO lines were observed. In the case of a double peak profile, we consider the two peaks at the same time to derive the total intensities. For the galaxies observed at IRAM 30m, the intensities were derived in units of the main beam temperature. To derive accurate line ratios, we must consider units of the brightness temperature and correct them from the beam difference. We derived the line ratio using $R_{21}=\frac{I_{CO21}\Omega_{CO21}}{I_{CO10}\Omega_{CO10}}$ (see \citealt{Wilson}). This equation assumes that in either line the sources are unresolved by the 30m beams. This is unlikely the case here (see below) but, in absence of spatial information on the CO emission, we use these line ratios as an indication.

   The line ratio for the galaxies in our sample is smaller than unity (see Table \ref{table:LCO-MH2}), which indicates that the CO emission is optically thick. If the gas was thermally excited, the excitation temperature should be smaller than 10 K to explain the line ratios (see \citealt{Braine_1992}). However, \cite{Ocana_2010, Husemann_2017,Shangguan_2020a} suggest that such low line ratios are most likely due to sub-thermal CO excitation.

   The line ratio presents a large scatter from galaxy-to-galaxy between 0.13 and 0.74. This is likely due to the fact that the CO emission is extended and, as expected, the hypothesis that the solid angle of the source is much smaller than the beam fails. In particular, increasing the size of the CO emission will increase the line ratio (e.g. \citealt{Combes_2011}). Moreover, the different fields of view between the lines can result in a large missing CO(2-1) emission.

   In the case of IRAS\,17020, the molecular gas content is fully covered by the CO(1-0) beam size. However, it extends further than the FWHM of the CO(2-1) beam size: $15''$ versus $11.6''$ (17.7 vs 13.7 kpc; \citealt{SalomeQ_2021}).
In 14 galaxies from the HERACLES survey, \cite{Leroy_2009} found that the CO emission extends to about half the radius of the B-band $25\: \rm mag\,arcsec^{-2}$ isophote. For IRAS\,13349, the molecular gas is thus expected to extend on the same scale as the FWHM of the beam of the CO(2-1). However, the distribution of the molecular gas in IRAS\,17020 extends to a radius about 15\% larger than half $r_{25}$ therefore, some CO(2-1) emission may have been missed.
In Ark\,564, Mrk\,1044, Mrk\,110 and Mrk\,766, the molecular gas is predicted to extend to about 80\% of the beam of the CO(1-0) emission. Therefore, a large fraction of the CO(2-1) emission is missing, implying that the line ratio is a lower limit.

   With the present observations, it is impossible to determine precisely the influence of the fields of view. To derive accurate line ratios and study the excitation of the CO emission in the galaxies of our sample, we therefore need to resolve the molecular gas in both lines with interferometric observations.
\medskip

   We cannot exclude that the CO(1-0) emission extends further than the CO(1-0) beam. However, the radial profile of the molecular gas in star-forming galaxies shows an exponential decrease \citep{Leroy_2009,Saintonge_2011}, indicating that the majority of the CO(1-0) emission has been recovered by our observations. \cite{Saintonge_2011} derived an aperture correction based on the optical diameter $D_{25}$ which we applied to our IRAM 30m galaxies. In the case of IRAS\,17020, we see that this can also correct the CO luminosity calculated using the formula of \cite{Solomon_1997} from the distribution of the emission within the beam. The corrected luminosities and molecular gas masses are reported in Table \ref{table:LCO-MH2}.


\begin{table*}
  \centering
  \caption{\label{table:incigale} Input parameter values used in the SED-fitting procedure.}
  \begin{tabular}{lccc}
    \hline \hline
    Template               &    Parameter      &                  Values                  & Description                                      \\ \hline
    {\it Stellar emission} &        IMF        &           \cite{Chabrier_2003}           & \\
                           &         Z         &                   0.02                   & Metallicity                                      \\
                           &  Separation age   &                  10 Myr                  & Separation age between the young                 \\
                           &                   &                                          & and the old stellar populations                  \\
    Delayed SFH            &     Age main      &    5.0, 7.0, 9.0, 11.0, 13.0$^*$ Gyr     & Age of the main SSP                              \\
                           &   $\tau_{main}$   &       0.10, 0.5, 1.0, 3.0, 5.0 Gyr       & e-folding time of the main SFH                   \\
                           &     Age burst     &  0.05, 0.10, 0.25, 0.50, 0.75, 1.0 Gyr   & Age of the late starburst                        \\
                           &  $\tau_{burst}$   &           25, 50, 75, 100 Myr            & e-folding time of the late starburst             \\
                           &    $f_{burst}$    &          0.01, 0.05, 0.1, 0.25           & Mass fraction of the late burst population       \\
    Modified Calzetti      &     $E(B-V)$      & 0.05, 0.1, 0.3, 0.5, 0.7, 0.9, 1.1, 1.3  & Attenuation of the                               \\
                           &                   &                                          & young stellar population                         \\
    Attenuation law        & Reduction factor  &                   0.44                   & Differential reddening applied to                \\
                           &                   &                                          & the old stellar population                       \\
                           &     $\delta$      &           $-0.6,-0.4,-0.2,0.0$           & Slope of the power law multiplying               \\
                           &                   &                                          & the Calzetti attenuation law                     \\ \hline
    {\it Dust emission}    &     $\alpha$      &       0.5, 1.0, 1.5, 2.0, 2.5, 3.0       & Slope of the power law combining                 \\
                           &                   &                                          & the contribution of different dust templates     \\ \hline
    {\it AGN emission}     & $R_{max}/R_{min}$ &               60, 100, 150               & Ratio of the outer and inner radii               \\
                           &   $\tau_{9.7}$    &               0.6, 3.0, 6.0              & Optical depth at $9.7\: \micron$                 \\
                           &      $\beta$      &              $-1.0,-0.5,0.0$             & Slope of the radial coordinate                   \\
                           &     $\gamma$      &                 0.0, 6.0                 & Exponent of the angular coordinate               \\
                           &     $\theta$      &               100, 140 deg               & Opening angle of the torus                       \\
                           &      $\psi$       &        40, 50, 60, 70, 80, 90 deg        & Angle between equatorial axis and line of sight  \\
                           &     $f_{AGN}$     &   0.3, 0.35, 0.4, 0.45, 0.5, 0.55, 0.6   & AGN fraction to the total IR emission            \\
                           &                   &  0.65, 0.7, 0.75, 0.8, 0.85, 0.9, 0.95   & \\ \hline
    {\it Nebular emission} &        $U$        &                 $10 ^{-2}$               & Ionization parameter                             \\
                           &     $f_{esc}$     &                    0\%                   & Fraction of Lyman continuum                      \\
                           &                   &                                          & photons escaping the galaxy                      \\
                           &    $f_{dust}$     &                   10\%                   & Fraction of Lyman continuum                      \\
                           &                   &                                          & photons absorbed by dust                         \\ \hline
  \end{tabular} \\
  \justify {\small
  \textbf{Notes.} 
    \tablefoottext{*}{For the galaxies at a redshift $z>0.04$, we use 12.0 Gyr as the maximum value.}
  }
\end{table*}


\begin{figure*}
  \centering
  \includegraphics[width=0.48\linewidth,trim=15 10 45 40,clip=true]{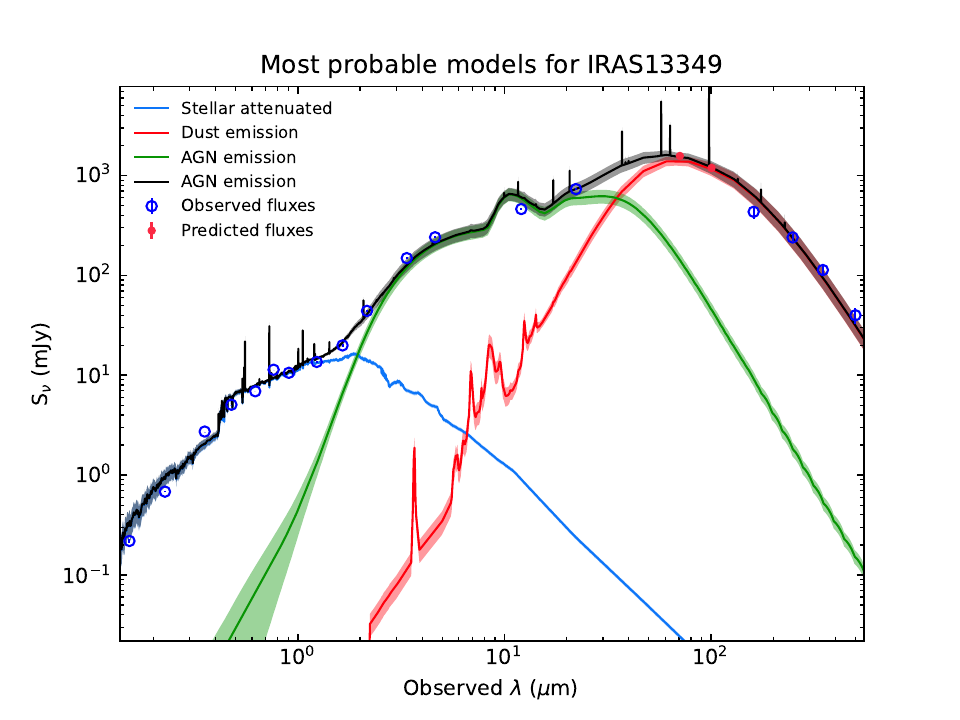}
  \hspace{3mm}
  \includegraphics[width=0.48\linewidth,trim=30 20 15 14,clip=true]{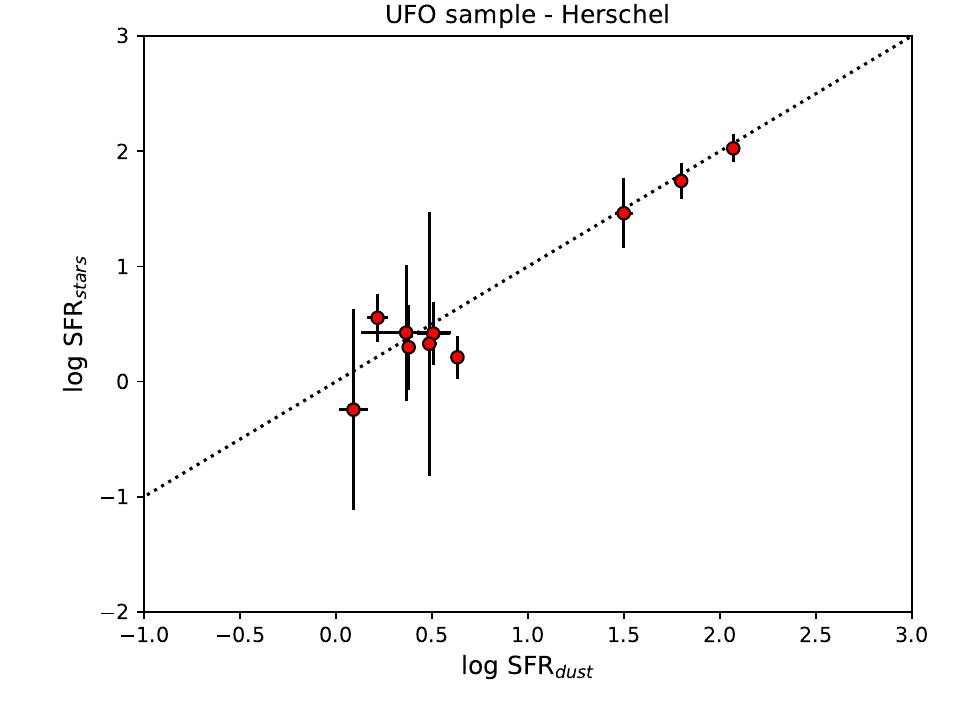}
  \caption{\label{cigale} \emph{Left:} Multi-band photometric spectral energy distributions (SED) of IRAS\,13349 as derived with CIGALE. The blue circles are the observed fluxes, while the red dots show the modeled fluxes for the Herschel/PACS bands with no observations/detection (see Section \ref{sec:cigale}, below). The blue, green and red shaded areas show the contribution of the stellar emission, the AGN emission and the dust thermal emission, respectively. Those shaded areas are compiled using the most probable input parameters returned by CIGALE and correspond to the galaxy properties reported in Table \ref{table:outcigale}.
  \emph{Right:} Comparison between the SFR derived by CIGALE from the stellar population modelling and the SFR derived from the dust FIR emission for the galaxies in our sample. The dotted line shows the identity relation.}
\end{figure*}


   \subsection{Multi-band SED fitting}
   \label{sec:cigale}

   To study star formation efficiency, we will compare the molecular gas mass reservoir as traced by the CO emission with global properties of the galaxies of our sample (e.g. stellar mass, star formation rate, AGN bolometric luminosities). To estimate these properties, we fitted the multi-band spectral energy distribution (SED) with the version 2020.0 of the Code Investigating GALaxy Emission (CIGALE; \citealt{CIGALE1,CIGALE2}).
The code considers several distinct emission components: (i) stellar emission, dominating the wavelength range $0.3-5\: \rm \mu m$; (ii) the FIR emission by cold dust; (iii) the emission from a central AGN, as direct energy coming from the accretion disc and reprocessed emission by the dusty torus; (iv) radio synchrotron emission. After assembling the models, given a range of input parameters, the code computes the model-expected fluxes that are compared to the observed photometry through a Bayesian statistical analysis.

   To build the SED, we compiled the multi-band photometry from the far ultraviolet (FUV) to the far infrared (FIR). The ultraviolet from GALEX \citep{GALEX} comes from the NED archive\footnote{\url{http://ned.ipac.caltech.edu/}}. In the optical, we used the photometry in the \emph{u',g',r',i',z'} filters of the SDSS\footnote{\url{https://skyserver.sdss.org/dr18/}} \citep{SDSS,SDSS-III}. For Mrk\,205, there is no ancillary data from SDSS therefore we used the photometry from the Hubble Source Catalog\footnote{\url{https://catalogs.mast.stsci.edu/hsc/}} (\citealt{Hubble_catalogue}; WFPC2 images with filters F439W, F555W, F814W).
The near- and mid-infrared photometry were taken from 2MASS (bands H, J, K$_S$; \citealt{2MASS}) and WISE (bands W1, W2, W3, W4; \citealt{WISE}), respectively.
Finally, for the far-infrared (FIR), we used Herschel/PACS and SPIRE data when available \citep{PACS,SPIRE}, otherwise we used IRAS in the 60 and $100\: \rm \mu m$ bands \citep{IRAS}. For I\,Zw\,1, PG\,1211 and PG\,1448, the Herschel photometry comes from \cite{Shangguan_2018}, while for IRAS\,13349 and Mrk\,110 we derived the fluxes following the method of \cite{Remy_2013}.
For Mrk\,205, we could only used the photometry from Spitzer/MIPS at $70\: \rm \mu m$ as there is no ancillary data from IRAS and Herschel.

   The stellar emission is modelled with the \cite{Bruzual_2003} stellar population models and a \cite{Chabrier_2003} initial mass function (IMF), assuming a delayed exponentially declining star formation history (SFH) for the main population and a decaying exponential for a late starburst event. We used a fixed metallicity $Z=0.02$, the solar value.
The modelled stellar emission is attenuated by dust, following a modified version of the \cite{Calzetti_2000} curve. A reduction factor $E(B-V)_{old}/E(B-V)_{young}=0.44$ \citep{Jarvis_2019} is applied to account a differential reddening for the old stellar population ($>10\: \rm Myr$).
The FIR dust emission is reproduced using the library of \cite{Dale_2014}, without contribution from the AGN. Therefore, the library only accounts for the contribution from star formation.
Finally, the AGN emission is treated separately using the torus models from \cite{Fritz_2006}. We only considered the AGN templates which are compatible with type 1 AGN ($\Psi\geq 40$\degree; see Table \ref{table:incigale}).

   For each set of parameters, CIGALE models the associated components of the SED. It can also derive the flux in given bands and estimate physical quantities, like the stellar mass, the star formation rate (SFR) or the AGN bolometric luminosity.
The code then uses a Bayesian-like approach by weighting the models depending on "their goodness-of-fit" \citep{CIGALE2}. This returns the most probable values for the output parameters, along with their uncertainties. We report the main physical quantities in Table \ref{table:outcigale} and show an example of SED fitting in Figure \ref{cigale} (left panel). The SEDs of the whole sample are presented in Appendix \ref{sec:SED}. When GALEX or Herschel data are missing, the flux in those band is modelled by CIGALE, which provides a prediction of the real flux for future observations. The associated errorbar is associated to the Bayesian-like approach.

   Two estimates of the SFR are reported in Table \ref{table:outcigale}: (1) one comes from the stellar population models and corresponds to the average star formation rate average over the last 100 Myr, (2) the other is derived from the dust FIR emission following \cite{Kennicutt_2012}. While the SFR calibration of \cite{Kennicutt_2012} is based on the \cite{Kroupa_2001} IMF, several studies report that the calibration based on the \cite{Chabrier_2003} IMF gives nearly identical results \citep{chomiuk_2011,Boissier_2013,Boselli_2015}. We see that the two estimates of the SFR are consistent with each other (Figure \ref{cigale} - right panel). In the following we will consider the SFR derived from the FIR emission, for consistency with the different samples we use as reference. For Mrk\,1044 and PG\,1448, this provides a much better estimate of the SFR than the stellar emission, which has relative uncertainties larger than unity.

\begin{figure}
  \centering
  \includegraphics[width=\linewidth,trim=25 15 45 40,clip=true]{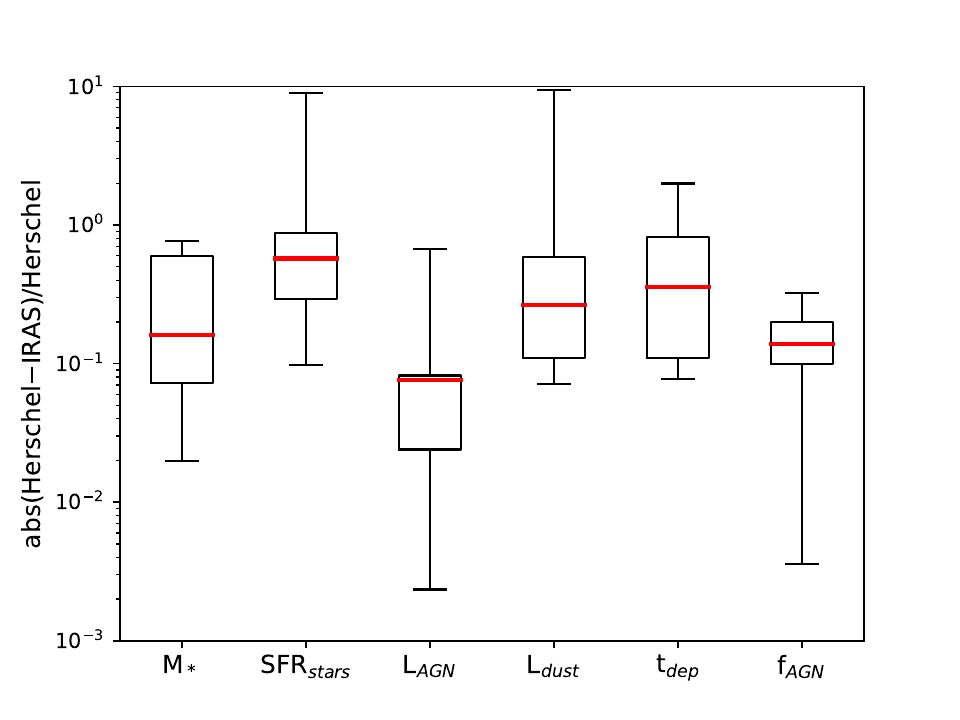}
  \caption{\label{Herschel} Influence of including the Herschel/PACS and SPIRE data. This box plot shows the relative difference between the values derived from the SED fitting CIGALE (1) when the Herschel/PACS and SPIRE data are used and (2) when only the IRAS data are used to model the FIR emission. The red lines and the boxes indicate the median values and the first and third quartiles, while the whiskers go from the minimum to the maximum value.}
\end{figure}

   Only six galaxies in our sample were observed with Herschel/PACS and SPIRE with bands at 70, 100, 160 $\rm \mu m$ and 250, 350, 500 $\rm \mu m$, respectively. For the other four sources, the FIR emission is only covered up to $70-100\: \rm \mu m$ by IRAS. However, the cold dust emission follows a modified black body which typically peaks around $100-160\: \rm \mu m$ therefore this component is only well constrained when the source is observed and detected by Herschel.
For the six galaxies with Herschel observations, we run the CIGALE code twice: (1) using the Herschel/PACS and SPIRE data to model the dust emission, and (2) limiting the observational constraints of the FIR emission to the IRAS data only. The properties computed by CIGALE typically varies by a factor up to 2-3, as indicating by the quartiles boxes. Nevertheless, we observed that in the case of PG\,1211, the SFR derived from the stellar emission and the dust emission vary by a factor of 10, highlighting the importance of the Herschel data. We thus conclude that some of the properties derived with CIGALE for IRAS\,17020, Mrk\,205, Ark\,564 and Mrk\,1044 may have been under/overestimated, in particular in the case of Mrk\,205 for which the predictions of the FUV and FIR fluxes deviate significantly from the SED fitting (Figure \ref{cigale-all}). However, with the current data, it is not possible to say whether this is the case or not.

\begin{table*}
  \centering
  \caption{\label{table:outcigale} Main properties of our sample galaxies as derived by CIGALE: (1) Short name; (2) Stellar mass of the host galaxy; (3) Average SFR over 100 Myr; (4) Bolometric luminosity of the AGN; (5) AGN-corrected FIR luminosity in the range $8-1000\: \rm \mu m$; (6) SFR derived from the dust emission following \cite{Kennicutt_2012}; (7) Molecular gas depletion time related to star formation; (8) Contribution of the AGN to the FIR luminosity.}
  \begin{tabular}{lccccccc}
    \hline \hline
    Galaxy      &         $M_*$         &   SFR ($<$100 Myr)  &        $L_{AGN}$         &          $L_{dust}$          &     $SFR_{FIR}$     & $t_{dep}^{mol}$ &   $f_{AGN}$    \\
                & ($10^{10}\: M_\odot$) & ($M_\odot.yr^{-1}$) & ($10^{44}\: erg.s^{-1}$) &         ($L_\odot$)          & ($M_\odot.yr^{-1}$) &      (Gyr)      &                \\ \hline
    IRAS\,13349 &    $21.44\pm 2.68$    &   $105.8\pm 29.5$   &      $99.7\pm 5.1$       & $(7.8\pm 0.4)\times 10^{11}$ &   $117.3\pm 5.9$    &      0.028      & $0.74\pm 0.02$ \\
    IRAS\,17020 &    $ 3.97\pm 1.44$    &   $ 29.0\pm 20.3$   &      $ 8.5\pm 0.9$       & $(2.1\pm 0.2)\times 10^{11}$ &   $ 31.5\pm 3.4$    &      0.038      & $0.43\pm 0.05$ \\
    Mrk\,205    &    $ 0.97\pm 1.44$    &   $  2.7\pm  3.6$   &      $ 7.4\pm 0.7$       & $(1.5\pm 0.8)\times 10^{10}$ &   $  2.3\pm 1.2$    &      1.854      & $0.85\pm 0.07$ \\
    Ark\,564    &    $ 0.93\pm 0.30$    &   $  2.6\pm  1.6$   &      $ 2.2\pm 0.3$       & $(2.1\pm 0.4)\times 10^{10}$ &   $  3.2\pm 0.6$    &      0.050      & $0.62\pm 0.06$ \\
    Mrk\,1044   &    $ 1.90\pm 0.55$    &   $  0.6\pm  1.1$   &      $ 1.1\pm 0.2$       & $(8.2\pm 1.4)\times 10^9$    &   $  1.2\pm 0.2$    &      0.527      & $0.51\pm 0.06$ \\
    Mrk\,110    &    $ 1.76\pm 0.66$    &   $  3.6\pm  1.7$   &      $ 1.4\pm 0.4$       & $(1.1\pm 0.1)\times 10^{10}$ &   $  1.6\pm 0.2$    &      0.468      & $0.72\pm 0.03$ \\
    I\,Zw\,1    &    $ 1.86\pm 0.89$    &   $ 55.2\pm 19.6$   &      $26.9\pm 1.6$       & $(4.2\pm 0.2)\times 10^{11}$ &   $ 62.8\pm 3.1$    &      0.070      & $0.54\pm 0.02$ \\
    PG\,1211    &    $ 0.58\pm 0.70$    &   $  2.0\pm  1.7$   &      $41.3\pm 2.1$       & $(1.6\pm 0.1)\times 10^{10}$ &   $  2.4\pm 0.1$    &      0.031      & $0.95\pm 0.00$ \\
    PG\,1448    &    $ 4.26\pm 1.53$    &   $  2.1\pm  5.6$   &      $ 9.0\pm 1.5$       & $(2.0\pm 0.2)\times 10^{10}$ &   $  3.1\pm 0.3$    &      0.523      & $0.80\pm 0.02$ \\
    Mrk\,766    &    $ 1.17\pm 0.20$    &   $  1.6\pm  0.7$   &      $ 0.7\pm 0.1$       & $(2.9\pm 0.1)\times 10^{10}$ &   $  4.3\pm 0.2$    &      0.070      & $0.28\pm 0.03$ \\
\hline
  \end{tabular}
\end{table*}

   Most of our sample galaxies have stellar masses $M_*\lesssim 4\times 10^{10}\: \rm M_\odot$. This is roughly compatible with the median value reported by \cite{Koutoulidis_2022} for their sample of type 1 galaxies, for which masses down to few $10^9\: \rm M_\odot$ are derived.
Only three galaxies in our sample have reported stellar masses in the literature: Mrk\,110 \citep{Bentz_2018}, I\,Zw\,1 and PG\,1448 \citep{Zhao_2021a}. The authors used a two-dimensional image decomposition with GALFIT and derived the stellar mass from the $V-H$ and $B-I$ color. For Mrk\,110 and PG\,1448, the stellar mass computed by CIGALE is consistent with the literature. On the contrary, for I\,Zw\,1 the stellar mass from CIGALE is a factor of about 7 lower than the mass reported by \cite{Zhao_2021a}.
For consistency with the rest of our sample, we decided to use the estimation from CIGALE. We note that the stellar mass is not the focus of this paper, neither is critical for the present analysis. In particular, in the following we use the SFR derived from the FIR luminosity.
Nevertheless, the stellar masses reported in Table \ref{table:outcigale} are model-dependent as a different set of input parameters may give a different mass. They should therefore be considered with caution in further studies and would need to be confirmed using an independent method.

   \subsection{Molecular gas mass}
   \label{sec:H2_mass}

   It is possible to derive the total molecular gas mass from the CO emission by applying a CO-to-H2 conversion factor $\alpha_{CO}=M_{H_2}/L'_{CO}$. The CO luminosity $L'_{CO}$ was calculated from the CO(1-0) emission using the formula of \cite{Solomon_1997}, then corrected for aperture effects (see Section \ref{sec:line_ratio} and Table \ref{table:LCO-MH2}).
There is a strong uncertainty about the value of this conversion factor (we refer to the review by \citealt{Bolatto_2013} for a complete discussion). While the conversion factor may vary by more than one order of magnitude at solar metallicity, typical values are commonly accepted for local galaxies.
For normal star-forming galaxies, \cite{Bolatto_2013} recommend to use a conversion factor of $4.3\: \rm M_\odot\,(K\,km\,s^{-1}\,pc^2)^{-1}$, which corresponds to the typical $X_{CO}$ observed in the Milky Way.
In starburst galaxies and LIRG/ULIRG, the molecular gas experiences very different conditions. The gas volume and column densities are much higher than in typical of normal disks and the molecular gas might not be virialized \citep{Downes_1998}, which results in a lower conversion factor. While the conversion factor presents a large dispersion, a value of $0.8\: \rm M_\odot\,(K\,km\,s^{-1}\,pc^2)^{-1}$ is commonly adopted in the literature (e.g. \citealt{Papadopoulos_2012,Cicone_2012, Cicone_2014,Veilleux_2017}).


\begin{figure}
  \centering
  \includegraphics[page=1,width=\linewidth,trim=25 15 10 10,clip=true]{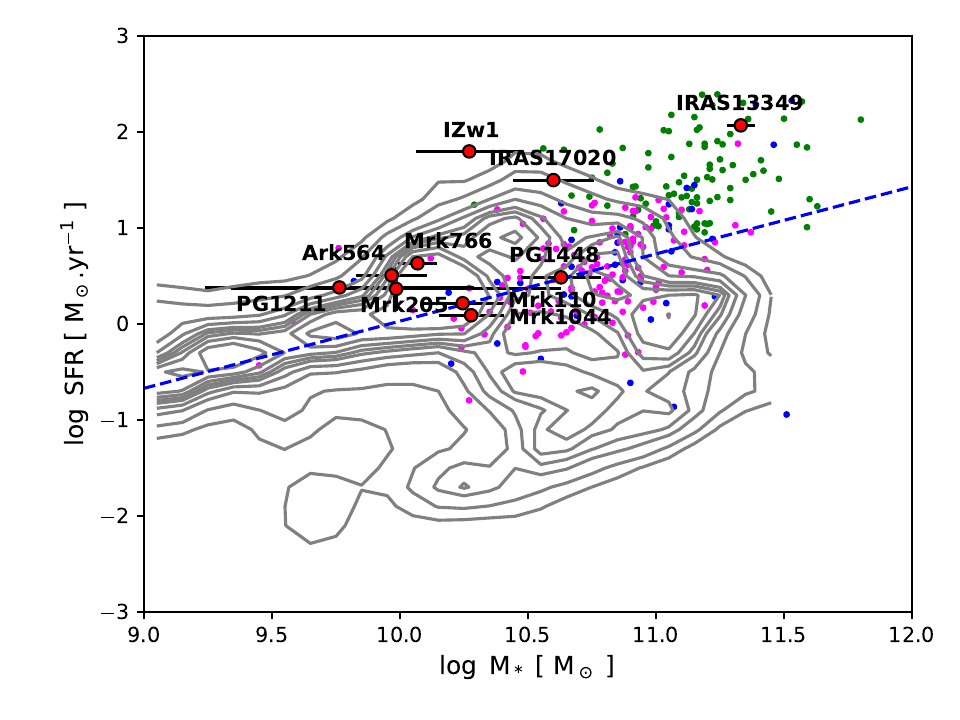}
  \caption{\label{SFMS} Position of the NLSy1 of our sample (red points with the corresponding error bars) relative to the main sequence of galaxies (SFR vs. $M_*$). The grey contours are those of the xCOLDGASS sample \citep{Saintonge_2011,Saintonge_2017}.
  The color points correspond to additional sample of galaxies with CO observations: PG quasars (blue, \citealt{Shangguan_2020a, Shangguan_2020b}), LIRG/ULIRG (green, \citealt{Gao_2004b,Gracia_2008,Armus_2009,Garcia_2012a,Tacconi_2018}) and BAT AGN (magenta, \citealt{Rosario_2018,Koss_2021}).
  The dashed line shows the definition of the MS from \cite{Whitaker_2012,Accurso_2017}.}
\end{figure}


   \cite{Sargent_2014} predicted variations of the $\alpha_{CO}$ factor with the stellar mass $M_*$ and SFR. The $M_*$-SFR plane shows the inhomogeneity of galaxy populations, in particular the bimodality between the blue star-forming galaxies and the red passive galaxies \citep{Baldry_2004,Daddi_2007, Elbaz_2007,Wuyts_2011}. According to \cite{Sargent_2014} the conversion factor suddently drops when the specific star formation rate $sSFR=SFR/M_*$ reaches about 3-4 times the value $sSFR_{MS}$ in the so-called "main sequence" of star-forming galaxies (hereafter MS). This provides a good observational diagnostic to separate starburst galaxies from "normal" star-forming galaxies.

   Figure \ref{SFMS} shows the position of the objects studied here in the $\mathrm{M_*}$-SFR plane, based on the SED fitting with CIGALE (see Sec. \ref{sec:cigale}). Six out of ten NLSy1 fall in the regime of starbursts, while Mrk\,110, Mrk\,1044 and PG\,1448 are located along the MS. The last galaxy (Mrk\,205) lie in between the two regimes. We therefore used a CO-to-H$_2$ conversion factor of $0.8\: M_\odot\,(K\,km\,s^{-1}\,pc^2)^{-1}$ for the six galaxies which lie high above the MS, and $4.3\: M_\odot\,(K\,km\,s^{-1}\,pc^2)^{-1}$ for Mrk\,110, Mrk\,205, Mrk\,1044 and PG\,1448.
The galaxies of our sample host a molecular gas mass between $7\times 10^7$ and $5\times 10^9\: \rm M_\odot$ (see Table \ref{table:LCO-MH2}).


\begin{figure*}
  \centering
  \includegraphics[page=2,height=6.6cm,trim=10 5 30 20,clip=true]{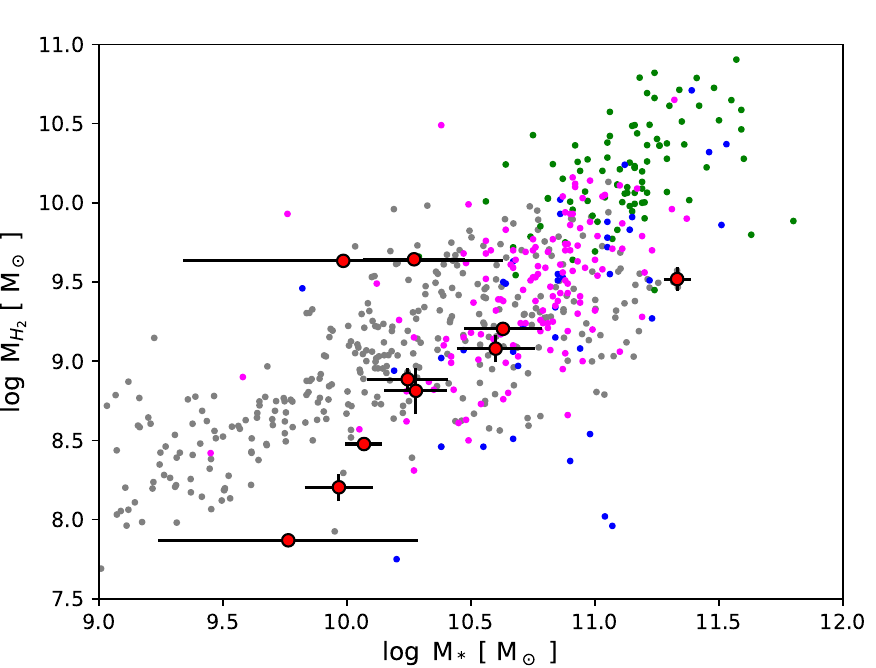}
  \hspace{3mm}
  \includegraphics[page=5,height=6.6cm,trim=20 5 30 20,clip=true]{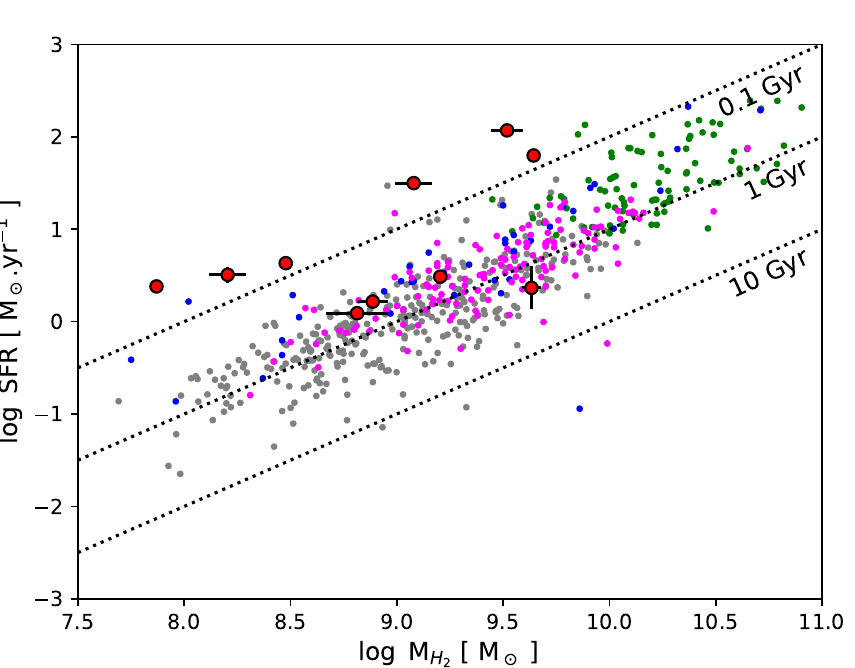}
  \caption{\label{CO-SFR} Molecular gas mass as a function of the stellar mass (\emph{left}) and SFR as a function of the molecular gas mass (\emph{right}). The colored symbols follow Figure \ref{SFMS}. The grey points are the CO detections of the xCOLDGASS sample \citep{Saintonge_2011,Saintonge_2017}. In the right panel, the dotted lines correspond to a depletion time of 0.1, 1, 10 Gyr (from top to bottom).}
\end{figure*}


   It is well established that the $\alpha_{CO}$ factor is also dependent on the gas metallicity: a lower metallicity means a lower CO abundance, resulting in a higher $\alpha_{CO}$ value \citep{Bolatto_2013,Accurso_2017}.
However, in absence of information on the metallicity in our sample, we do not take into account the metallicity-dependence of the $\alpha_{CO}$ factor. Integral field unit (IFU) observations of optical emission lines can provide information about the metal content in the host galaxies but this is out the scope of the present paper.

\section{Molecular gas content and star formation}
\label{sec:gas-SF}

   To put our NLSy1 sample in the larger context of galaxy evolution, we compare it with other samples of nearby ($z<0.1$) galaxies. (1) The xCOLDGASS survey \citep{Saintonge_2011,Saintonge_2017} observed 532 galaxies at redshift $0.025<z<0.05$ in CO(1-0) with the IRAM 30m telescope. The survey is a compilation of two sub-samples which covered different ranges of stellar mass: $log(M_*/M_\odot)=10-11.5$ for COLDGASS \citep{Saintonge_2011} and $log(M_*/M_\odot)=9-10$ for COLDGASS-low \citep{Saintonge_2017}.
(2) \cite{Tacconi_2018} compiled a sample of 90 luminous and ultraluminous infrared galaxies (LIRGs and ULIRGs) at $z=0.002-0.09$ with CO(1-0) detections from the literature \citep{Gao_2004b,Gracia_2008,Armus_2009,Garcia_2012a}.
(3) \cite{Shangguan_2020a,Shangguan_2020b} conducted ACA observations of the CO(2-1) line for a sample of 23 Palomar-Green quasars at $z<0.1$.
(4) \cite{Rosario_2018} and \cite{Koss_2021} present a compilation of CO observations from the Atacama Pathfinder EXperiment (APEX) and the James Clerk Maxwell Telescope (JCMT) for a sample of 233 AGN host galaxies at $z<0.05$, selected from the \emph{Swift}-BAT all sky catalogue.
\medskip

\noindent \textit{Position relative to the MS} - 
As stated in Section \ref{sec:H2_mass}, six out of ten NLSy1 fall in the regime of starbursts, while three are located along the MS (Figure \ref{SFMS}). The galaxies in our sample are in agreement with previous studies who concluded that AGN hosts mostly lie on or above the MS \citep{Mullaney_2015,Kakkad_2017,Koss_2021, Koutoulidis_2022}.
\medskip

\noindent \textit{Molecular gas fraction} - Several studies comparing the molecular gas fraction between active and inactive galaxies present opposite conclusions, with the active galaxies being more gas-rich \citep{Scoville_2003, Bertram_2007,Vito_2014,Koss_2021} or, on the contrary, containing less molecular gas \citep{Fiore_2017,Kakkad_2017} than inactive galaxies. Finally, \cite{Maiolino_1997} and \cite{Rosario_2018} found no difference in the molecular gas mass between active and inactive galaxies, at a given stellar mass. In the literature, various definitions are used for the molecular gas fraction. Here, we adopt the convention $f_{H_2}=M_{H_2}/M_*$.
The NLSy1 of our sample are in agreement with the general trend of the increasing molecular gas mass with increasing stellar mass (Figure \ref{CO-SFR} - left). However, they tend to present small molecular gas fractions in comparison to the other surveys described above: $f_{H_2}\sim 0.02-0.04$, except for Mrk\,205 ($f_{H_2}\sim 0.45$) and I\,Zw\,1 ($f_{H_2}\sim 0.24$).
\medskip

\noindent \textit{Star formation efficiency} - To study the global star formation efficiency in galaxies, the most commonly used method is the Kennicutt-Schmidt relation \citep{Kennicutt_1998a} between the surface densities of the molecular gas mass and the star formation rate. However we have no information about the star formation distribution and most of the molecular gas is not or barely resolved. As an alternative, we therefore plotted the relation between the molecular gas mass $M_{H_2}$ and the SFR (Figure \ref{CO-SFR} - right).
Except Mrk\,205, the NLSy1 galaxies in our sample lie above the relation of the star-forming galaxies with molecular depletion times down to a few 10 Myr, 1-2 order of magnitude shorter than the typical $\sim 2\: \rm Gyr$ in star-forming disc \citep{Leroy_2013}. Previous analysis comparing active and non active galaxies also found that AGN host galaxies are more efficient in forming stars \citep{Maiolino_1997,Fiore_2017,Kakkad_2017,Shangguan_2020b}. On the contrary, \cite{Rosario_2018} and \cite{Koss_2021} did not find any evidence that the star formation efficiency is affected by the presence of a central AGN.


\begin{figure*}
  \centering
  \includegraphics[page=2,height=6.4cm,trim=20 15 10 10,clip=true]{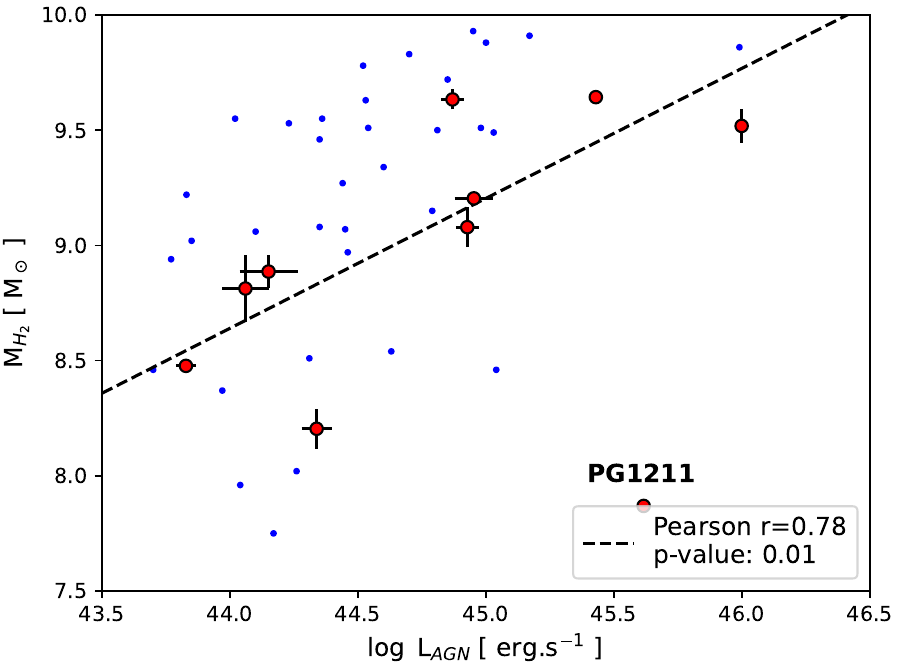}
  \hspace{2mm}
  \includegraphics[page=2,height=6.4cm,trim=20 15 10 10,clip=true]{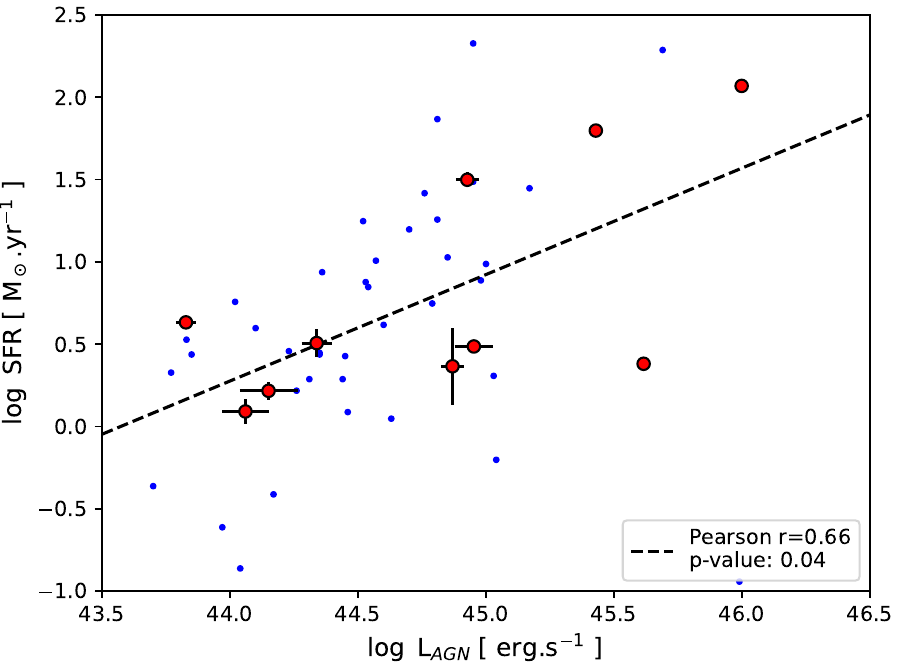} \\
  \vspace{2mm}
  \includegraphics[page=4,height=6.4cm,trim=20 15 10 10,clip=true]{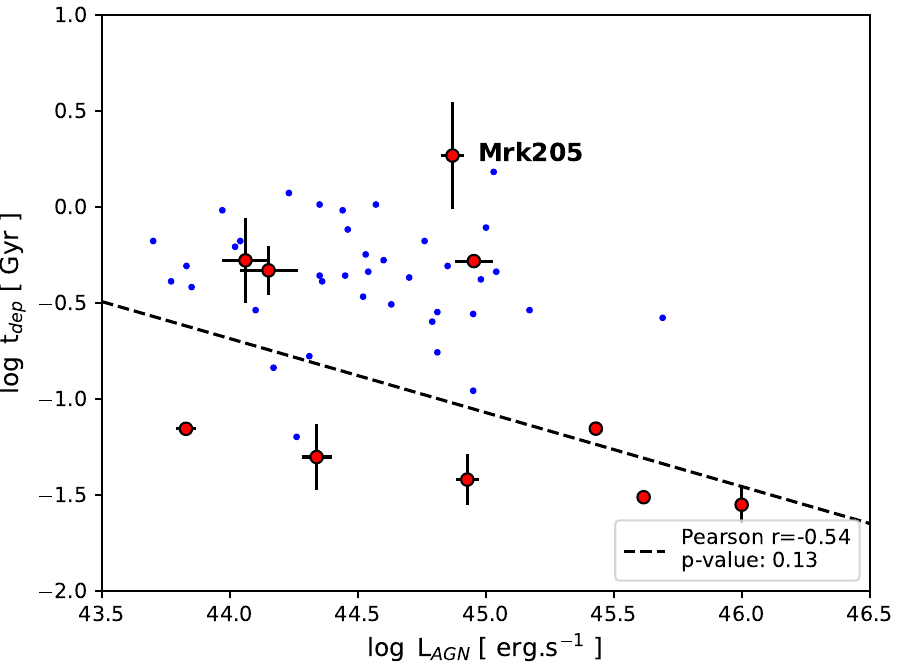}
  \hspace{2mm}
  \includegraphics[page=7,height=6.4cm,trim=20 15 10 10,clip=true]{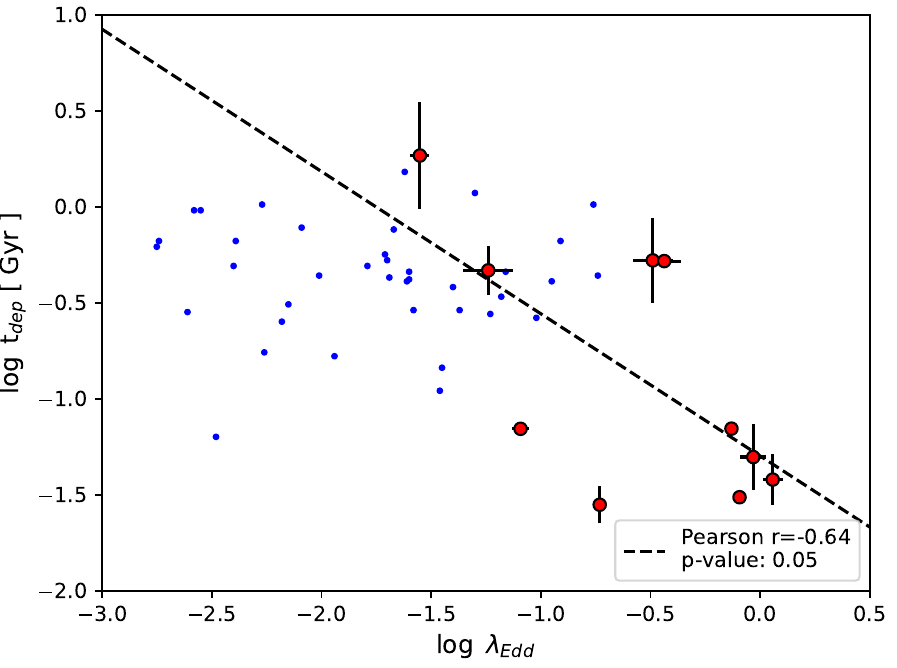}
  \caption{\label{gas-SFR-AGN} \emph{Top:} Molecular gas mass (\emph{left}) and star formation rate (\emph{right}) as a function of the AGN bolometric luminosity. Both quantities are correlated with the AGN luminosity (Pearson $r=0.7-0.8$; p-value$\,\leq 0.04$).
  \emph{Bottom:} Molecular gas depletion time as a function of the AGN luminosity (\emph{left}) and the Eddington ratio $\lambda_{Edd}=L_{Edd}/L_{AGN}$. We observe a tentative anti-correlation with the AGN luminosity when not considering Mrk\,205 (Pearson $r=-0.54$; p-value$\,\sim 0.13$) and a clear anti-correlation with the Eddington ratio (Pearson $r=-0.64$; p-value$\,\sim 0.05$). The blue points correspond to the PG quasars from \cite{Shangguan_2020a,Shangguan_2020b}.}
\end{figure*}


\section{Impact of the AGN activity}
\label{sec:AGN-feedback}

   In this section, we discuss the impact of the AGN activity on the molecular gas reservoir and star formation in the galaxies in our sample. In the following, all the Pearson's coefficient $r$ and the p-values are derived for our sample of 10 galaxies only. We observe that the CO luminosity and the molecular gas mass correlate well with the AGN bolometric luminosities (Pearson $r=0.78-0.87$, p-value$\,\leq 0.01$; Figure \ref{gas-SFR-AGN} - top left) when excluding PG\,1211 which appears as an outlier. Note that the correlation is less clear when PG\,1211 is considered ($r=0.33-0.55$, p-value of $0.1-0.4$).
Conversely, the molecular gas mass fraction $M_{H_2}/M_*$ is not correlated with the AGN luminosity (Pearson's coefficient $r=0.17$; p-value: 0.67).
Note that we observe no correlation between the molecular gas reservoir and the Eddington ratio $\lambda_{Edd}=\frac{L_{AGN}}{L_{Edd}}$, contrary to \cite{Koss_2021} who report a correlation for the gas mass fraction $M_{H_2}/M_*$.

   The SFR shows a clear correlation with the AGN luminosity ($r=0.66$, p-value of 0.04; Figure \ref{gas-SFR-AGN} - top right). This suggests that the SFR could be regulated by the AGN activity and not by the molecular gas reservoir, contrary to the conclusion of \cite{Baker_2022} for 46 local non-active galaxies. Moreover, we observe a correlation of the star formation efficiency $SFE=SFR/M_{H_2}=1/t_{dep}^{mol}$ with the Eddington ratio (Pearson $r=0.64$; p-value$\sim 0.05$; Figure \ref{gas-SFR-AGN} - lower right), as well as a tentative correlation with the AGN luminosity (Pearson $r=0.54$; p-value$\sim 0.13$; Figure \ref{gas-SFR-AGN} - lower left) if we do not consider Mrk\,205 for which the SFE is not higher than star-forming galaxies. The results differ from \cite{Shangguan_2020b} who found no dependence in their sample.

   \cite{Bischetti_2021} studied the cold gas properties within the host-galaxies of a sample of 8 high-redshift quasi-stellar objects (QSO) and observed both low molecular gas fractions and short depletion times which they interpreted as the result of AGN feedback. The depletion time in our sample of NLSy1 is of the same order of that observed by \cite{Bischetti_2021}, suggesting that AGN feedback might be at work in the sources of our sample.
This hypothesis is also supported by the nuclear wind activity that characterizes all of them, and by the recent findings reported in \cite{Longinotti_2023}, where energy-conservation was confirmed for the galaxy scale molecular outflow and the X-ray UFO.
In the following, we explore different scenarios which could explain the low depletion times we observe. Nevertheless, the current observations and ancillary data do not enable to disentangle between the different scenarios.
\medskip

\noindent \textit{Positive AGN feedback?} - 
AGN feedback is often invoked as a quenching mechanism (negative feedback). However, there is evidence of AGN positive feedback which favour or trigger star formation (e.g. \citealt{Zinn_2013}). While positive feedback is mostly observed by the presence of recent star formation along radio jets \citep{Klamer_2004, Emonts_2014,SalomeQ_2015a,SalomeQ_2016b,Zovaro_2020}, enhanced star formation is also observed around cavities produced by AGN-driven outflows \citep{Cresci_2015a, Cresci_2015b,Cresci_2023}.
Therefore, the AGN activity in our sample might be boosting star formation by compressing the gas. The $M_{H_2}-L_{AGN}$ correlation suggests that the AGN activity might also regulate the molecular gas reservoir and fulfill the observed starburst episode.
In this scenario, the short depletion times indicate that the AGN activity will quench star formation by gas exhaustion.
\medskip

\noindent \textit{Weak or no AGN feedback?} - 
Several studies proposed a scenario  where the AGN activity is ignited by a supply of gas from stellar feedback and supernovae explosions in the central region \citep{Chen_2009,Dahmer_2022,Tillman_2022}. Moreover, recent observations and simulations favour a scenario where AGN feedback is delayed (see \citealt{Cresci_2018} and references therein). In this scenario, the observed starburst event may have ignited the AGN activity without having a major impact on the star formation in the host of the NLSy1.

   As to the effect of the morphology of the  galaxies on their SFR, we note that in a sample of Seyfert 2 galaxies, \cite{Maiolino_1997} concluded that the high SFR they observed is the result of morphology perturbations. \cite{Krongold_2002} proposed an evolutionary sequence between galaxy interactions, starbursts episodes and AGN activity. While NLSy1 mostly reside in late-type galaxies \citep{Krongold_2001,Ohta_2007}, some are found in disturbed or interacting systems \citep{Jarvela_2018,SalomeQ_2021}.
Another process which has the potential to ignite the nuclear activity is the presence of stellar bars \citep{Simkin_1980,Shlosman_1990,Maiolino_2000,Garcia_2012b}. In particular, \cite{Crenshaw_2003} found that the presence of a large-scale stellar bar is common in NLSy1.

   A detailed study of the morphology of the galaxies in our sample is out of the scope of this paper. However, we performed a visual inspection of SDSS and HST images. A stellar bar is clearly present in at least half of our sample. Conversely, only two galaxies show evidence of galaxy interaction: Mrk\,110 with the presence of large-scale tidal tails, and I\,Zw\,1 \citep{Scharwachter_2003}. This suggests that the nuclear activity in our sample is regulated by the molecular gas reservoir which fuels the nucleus via the stellar bar. Nevertheless, the presence of galaxy interaction in more sources cannot be excluded. For example, using resolved CO observations, \cite{SalomeQ_2021} discovered that the host galaxy of IRAS\,17020 is interacting with a smaller mass companion, while the optical images do not show any evidence of interaction.
\medskip

\noindent \textit{Early phase of AGN feedback?} - 
\cite{Garcia_2021} observed the molecular gas in the centre of 13 Seyfert galaxies at a resolution of $\sim 7-10\: pc$ with ALMA. These observations allowed them to resolve the dusty molecular tori which extend to diameters $\leq 100\: pc$ and reveal clear evidence of AGN feedback at small scales. In particular, \cite{Garcia_2021} highlighted that the imprint of AGN feedback at small scales is likely to be more extreme in higher luminosity and/or higher Eddington ratio Seyfert galaxies, contrary to what we observe at larger scales in this paper.
However, AGN activity occurs on rather short timescales (few Myr; \citealt{Martini_2004,Hickox_2014} and NLSy1 have been dubbed as a class of young AGN in terms of the black hole mass (see Section \ref{sec:introduction}). Moreover numerical simulations predict that the super-Eddington accretion is short \citep{Li_2012,Smith_2019a} and does not affect star formation on short timescales \citep{Massonneau_2023}. All together, this suggests that the current AGN activity we observe in our sample may not have had time yet to affect galaxy-scale properties like the mass of stars and gas or the SFR, which extend to several kpc. This is in agreement with the scenario of \cite{Cresci_2018} where the so-called "negative feedback" does not affect the full gas reservoir but is rather delayed.

\section{Conclusion}
\label{sec:conclusion}

   In this paper, we studied the molecular gas reservoir of a sample of NLSy1 for which a relativistic UFO has been detected in X-ray. Six galaxies were observed in CO with the IRAM 30m telescope, including the galaxy IRAS\,17020 which was previously observed with NOEMA \citep{SalomeQ_2021,Longinotti_2023}. We complemented our sample with four NLSy1 detected in CO by NOEMA or ACA \citep{Cicone_2014,Shangguan_2020a,Dominguez-Fernandez_2020}.
We compared the molecular gas mass estimated from the CO observations with properties of the host galaxies and the AGN which were derived with the multi-band SED fitting code CIGALE.

   The CO molecule is detected in the ten galaxies considered in this article. Seven sources were observed in CO(1-0) and CO(2-1), and five are detected in both CO lines. The intensity line ratio $I_{CO21}/I_{CO10}$ suggests that the lines are not thermalised. We used the commonly adopted $\alpha_{CO}$ values based on the position of the galaxies in the $M_*-SFR$ plane (see Section \ref{sec:H2_mass}) and we derived molecular gas masses covering almost two orders of magnitude between $7\times 10^7$ and $4\times 10^9\: \rm M_\odot$, typical of the molecular gas mass observed in local galaxies by \cite{Saintonge_2017}.

   By fitting the multi-band SED with CIGALE, we computed the stellar mass and star formation rate of the host galaxies, as well as the AGN bolometric and the dust FIR luminosities. These properties also spans two orders of magnitude. The NLSy1 in our samples tend to host less molecular gas that local galaxies of the same $M_*$.
However, the star formation rate is significantly higher, resulting in molecular depletion times 1-2 orders of magnitude shorter than in typical star-forming galaxies (except in Mrk\,205). Moreover, we found that the SFR increases with increasing AGN luminosity and the molecular depletion time is anticorrelated with the AGN luminosity and the Eddington ratio. This suggests that the AGN activity has little or no quenching feedback on star formation and conversely might be enhancing star formation (positive feedback?).

   To further investigate the possible relation between the molecular gas content and the AGN activity, additional observations would be necessary. Firstly, interferometric observations enable to resolve the CO emission and study the distribution of the molecular gas. In particular, NOEMA data enabled \cite{SalomeQ_2021} to discover that IRAS\,17020 is currently interacting with a small companion. This interaction was not known before that as the host galaxy is observed as a typical undisturbed spiral barred galaxy in optical images \citep{Ohta_2007,Olguin_2020}. Secondly, H\rmnum{1} observations will trace the atomic phase of the ISM and enable to get a more complete view of the gas reservoir. Finally, CO observations of a larger sample of NLSy1 will provide more statistic to resolve the impact of the AGN activity in NLSy1. For instance, the IBISCO survey contains 8 NLSy1 \citep{Molina_2018} with unpublished CO observations \citep{Feruglio_2018}, including some ALMA data (mentionned by \citealt{Zanchettin_2021}).

\section*{Acknowledgements}

We thank the referee for the helpful and constructive comments.

Q.S. thanks Denis Burgarella for his help with CIGALE, and Jonathan Freundlich for sharing the Tacconi et al.'s catalogue.

This work is based on observations carried out under project number 065-20 with the IRAM 30m telescope. IRAM is supported by INSU/CNRS (France), MPG (Germany) and IGN (Spain).

The computer resources of the Finnish IT Center for Science (CSC) and the FGCI project (Finland) are acknowledged.

This research has made use of the NASA/IPAC Extragalactic Database (NED), which is operated by the Jet Propulsion Laboratory, California Institute of Technology, under contract with the National Aeronautics and Space Administration.

Funding for the SDSS and SDSS-II has been provided by the Alfred P. Sloan Foundation, the Participating Institutions, the National Science Foundation, the U.S. Department of Energy, the National Aeronautics and Space Administration, the Japanese Monbukagakusho, the Max Planck Society, and the Higher Education Funding Council for England. The SDSS Web Site is http://www.sdss.org/.

The SDSS is managed by the Astrophysical Research Consortium for the Participating Institutions. The Participating Institutions are the American Museum of Natural History, Astrophysical Institute Potsdam, University of Basel, University of Cambridge, Case Western Reserve University, University of Chicago, Drexel University, Fermilab, the Institute for Advanced Study, the Japan Participation Group, Johns Hopkins University, the Joint Institute for Nuclear Astrophysics, the Kavli Institute for Particle Astrophysics and Cosmology, the Korean Scientist Group, the Chinese Academy of Sciences (LAMOST), Los Alamos National Laboratory, the Max-Planck-Institute for Astronomy (MPIA), the Max-Planck-Institute for Astrophysics (MPA), New Mexico State University, Ohio State University, University of Pittsburgh, University of Portsmouth, Princeton University, the United States Naval Observatory, and the University of Washington.

Funding for SDSS-III has been provided by the Alfred P. Sloan Foundation, the Participating Institutions, the National Science Foundation, and the U.S. Department of Energy Office of Science. The SDSS-III web site is http://www.sdss3.org/.

SDSS-III is managed by the Astrophysical Research Consortium for the Participating Institutions of the SDSS-III Collaboration including the University of Arizona, the Brazilian Participation Group, Brookhaven National Laboratory, Carnegie Mellon University, University of Florida, the French Participation Group, the German Participation Group, Harvard University, the Instituto de Astrofisica de Canarias, the Michigan State/Notre Dame/JINA Participation Group, Johns Hopkins University, Lawrence Berkeley National Laboratory, Max Planck Institute for Astrophysics, Max Planck Institute for Extraterrestrial Physics, New Mexico State University, New York University, Ohio State University, Pennsylvania State University, University of Portsmouth, Princeton University, the Spanish Participation Group, University of Tokyo, University of Utah, Vanderbilt University, University of Virginia, University of Washington, and Yale University.

Based on observations made with the NASA/ESA Hubble Space Telescope, and obtained from the Hubble Legacy Archive, which is a collaboration between the Space Telescope Science Institute (STScI/NASA), the Space Telescope European Coordinating Facility (ST-ECF/ESAC/ESA) and the Canadian Astronomy Data Centre (CADC/NRC/CSA).

This publication makes use of data products from the Two Micron All Sky Survey, which is a joint project of the University of Massachusetts and the Infrared Processing and Analysis Center/California Institute of Technology, funded by the National Aeronautics and Space Administration and the National Science Foundation.

This publication makes use of data products from the Wide-field Infrared Survey Explorer, which is a joint project of the University of California, Los Angeles, and the Jet Propulsion Laboratory/California Institute of Technology, funded by the National Aeronautics and Space Administration.

Herschel is an ESA space observatory with science instruments provided by European-led Principal Investigator consortia and with important participation from NASA.

Q.S. and A.L.L. acknowledge support from CONACyT research grants CB-2016-01-286316. A.L.L. also acknowledges support from DGAPA-PAPIIT grant IA101623.

Y.K. acknowledges support from DGAPA-PAPIIT grant IN102023.

M.B. acknowledges support from the PRIN MIUR project "Black Hole winds and the Baryon Life Cycle of Galaxies: the stone-guest at the galaxy evolution supper", contract number 2017PH3WAT

S.G.B. acknowledges support from the research project PID2019-106027GA-C44 of the Spanish Ministerio de Ciencia e Innovaci\'on.

\section*{Data availability}

The IRAM 30m data underlying this article will be shared on reasonable request to the corresponding author.

\newcommand{\newblock}{}
\bibliography{Biblio}
\bibliographystyle{mnras}


\appendix

\section{CO spectra from IRAM 30m}
\label{sec:spectra}

\begin{figure*}
  \centering
  \includegraphics[width=0.48\linewidth,trim=30 40 0 40,clip=true]{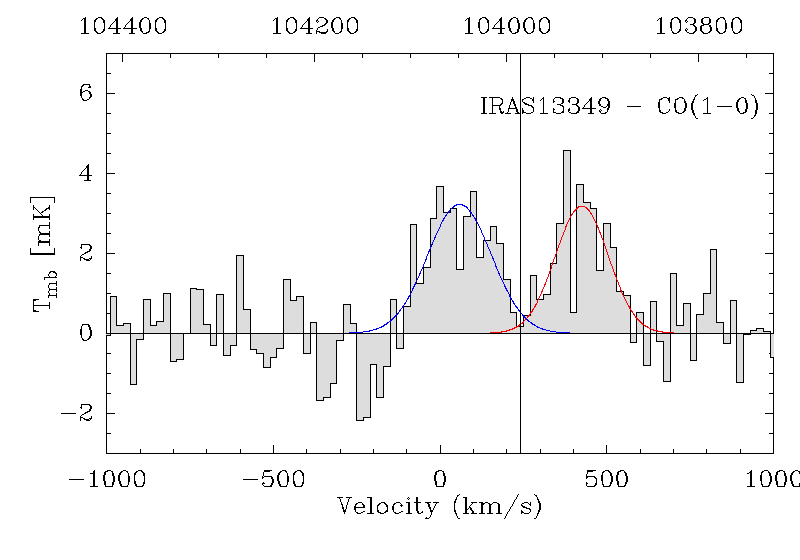}
  \hspace{3mm}
  \includegraphics[width=0.48\linewidth,trim=30 40 0 40,clip=true]{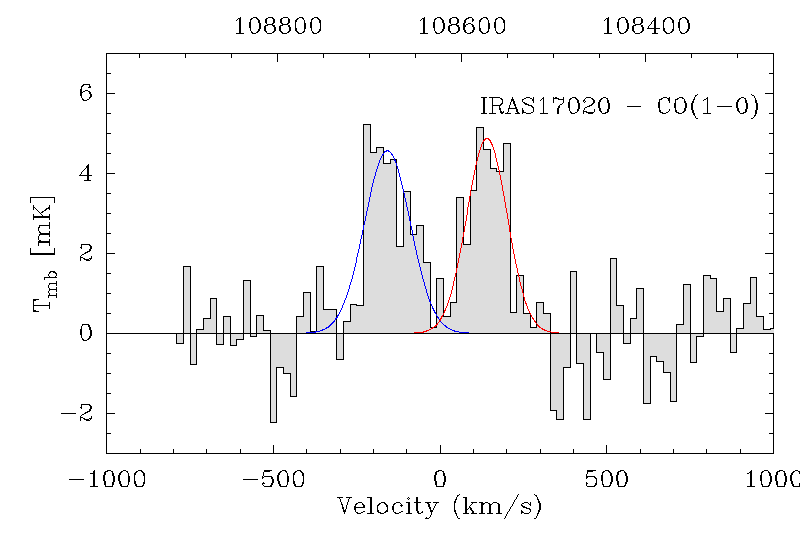} \\
  \vspace{3mm}
  \includegraphics[width=0.48\linewidth,trim=30 40 0 40,clip=true]{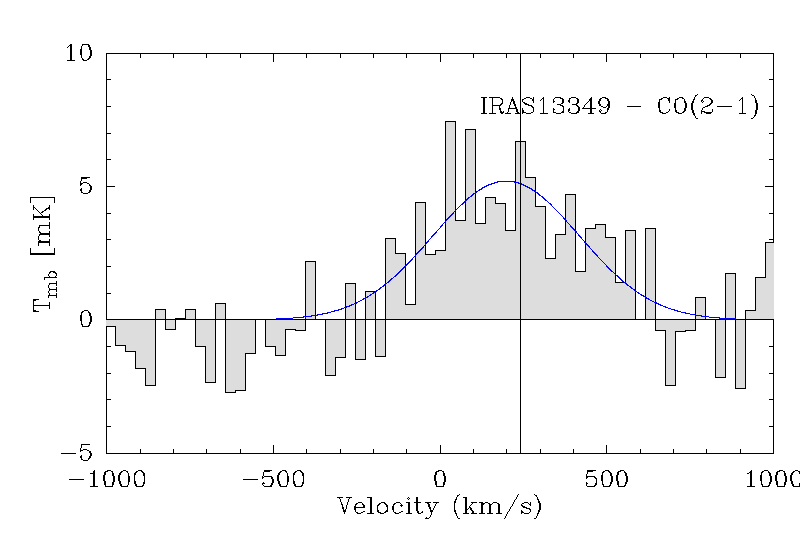}
  \hspace{3mm}
  \includegraphics[width=0.48\linewidth,trim=30 40 0 40,clip=true]{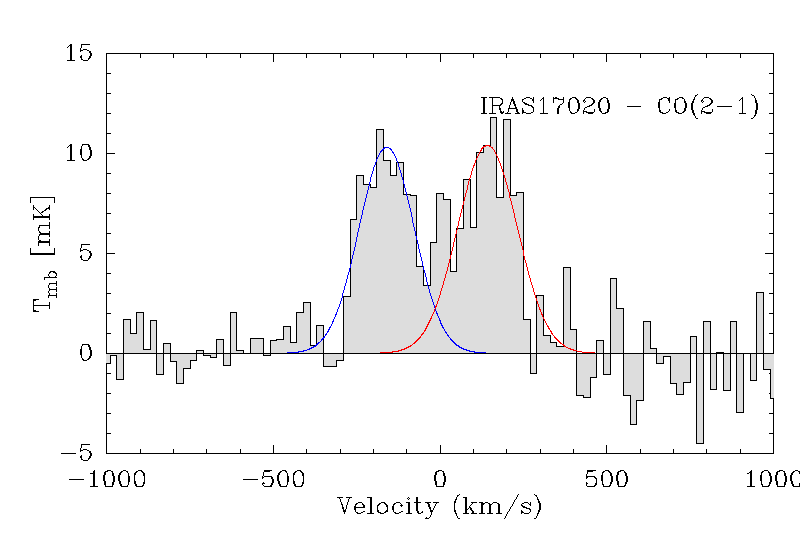} \\
  \vspace{3mm}
  \includegraphics[width=0.48\linewidth,trim=30 40 0 40,clip=true]{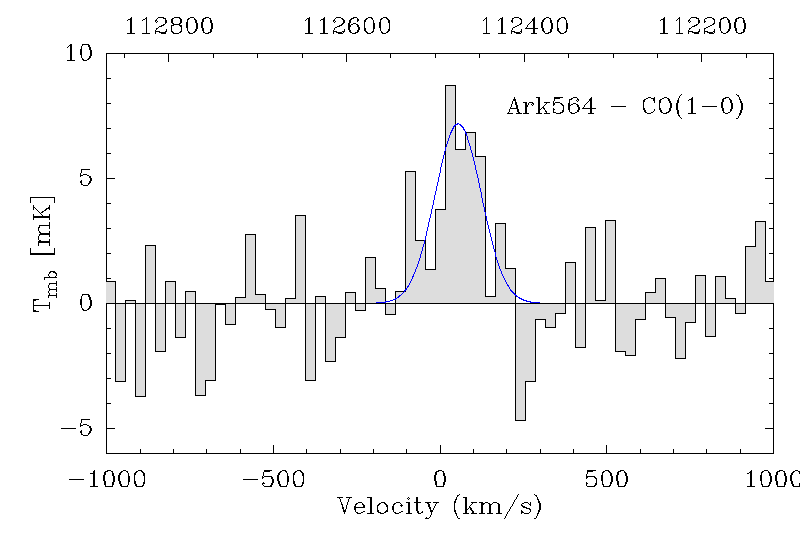}
  \hspace{3mm}
  \includegraphics[width=0.48\linewidth,trim=30 40 0 40,clip=true]{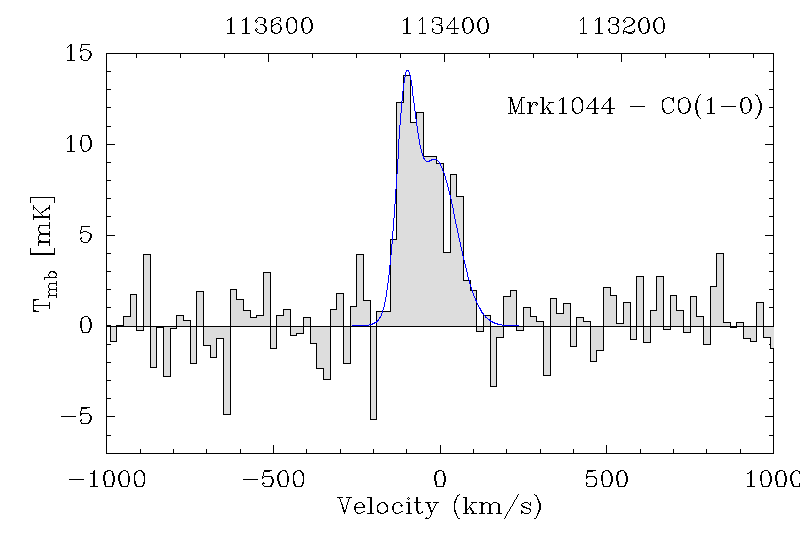} \\
  \vspace{3mm}
  \includegraphics[width=0.48\linewidth,trim=30 40 0 40,clip=true]{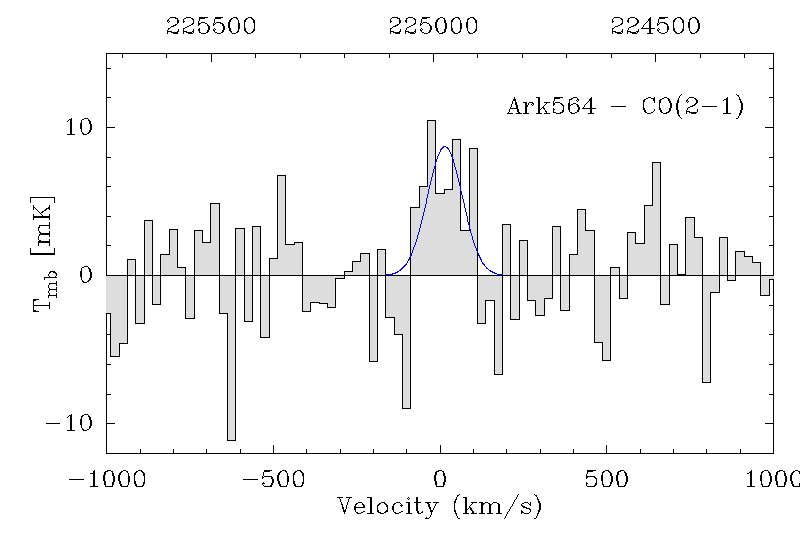}
  \hspace{3mm}
  \includegraphics[width=0.48\linewidth,trim=30 40 0 40,clip=true]{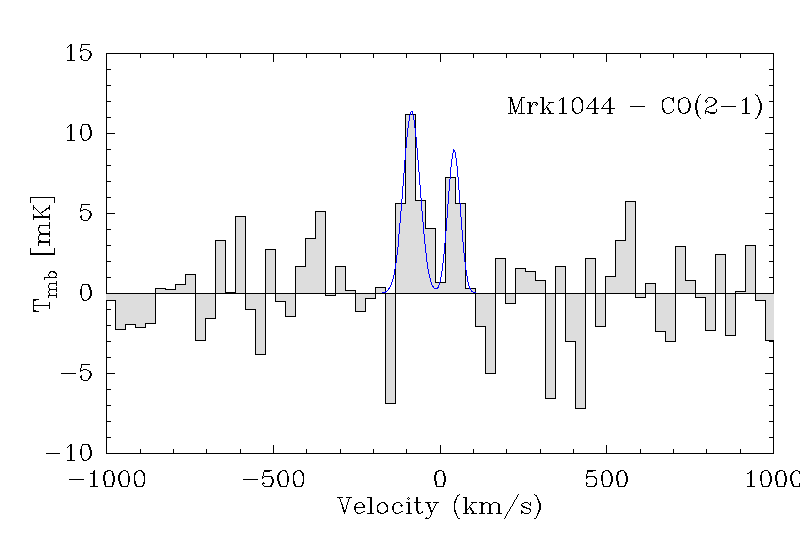}
  \caption{\label{spectra} Spectra of the CO(1-0) and CO(2-1) emission as observed with the IRAM 30m. For Mrk\,205, it was not possible to simultaneously observe both CO(1-0) and CO(2-1). The vertical line in the spectra of IRAS\,13349 correspond to our new estimate of the redshift.}
\end{figure*}

\begin{figure*}
  \centering
  \includegraphics[width=0.48\linewidth,trim=30 40 0 40,clip=true]{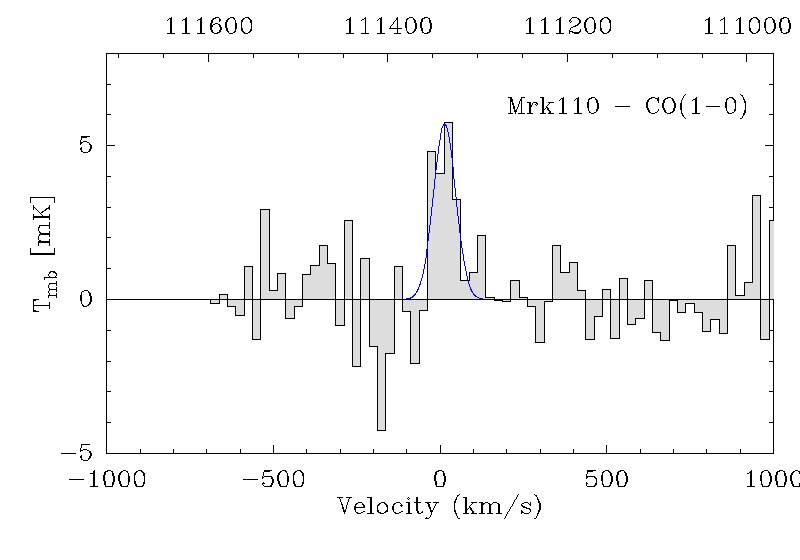}
  \hspace{3mm}
  \includegraphics[width=0.48\linewidth,trim=30 40 0 40,clip=true]{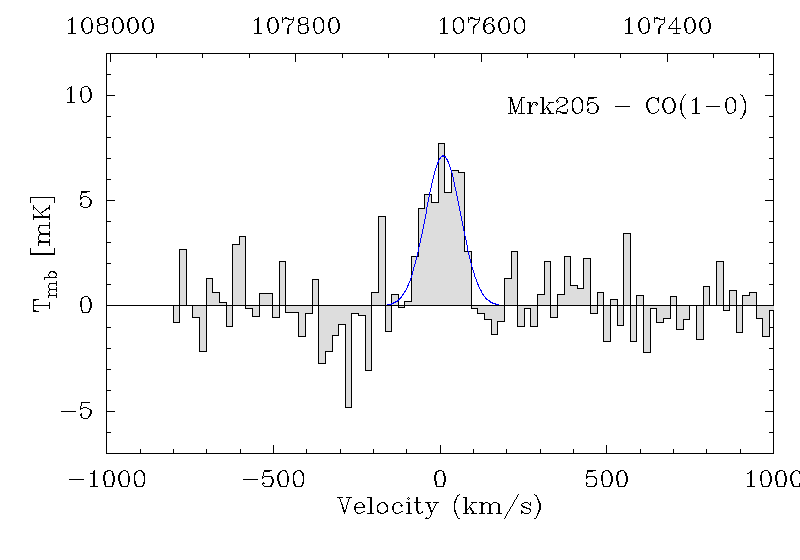} \\
  \vspace{3mm}
  \begin{justify}
    \includegraphics[width=0.48\linewidth,trim=30 40 0 40,clip=true]{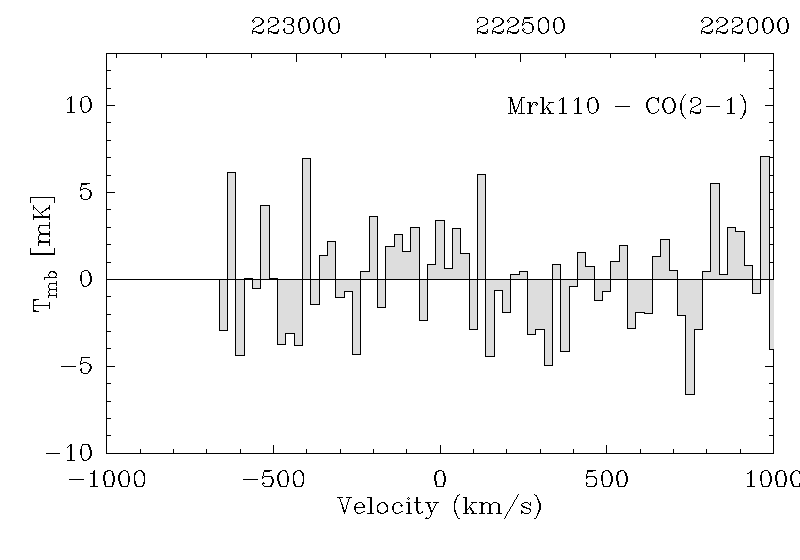} \\
  \end{justify}
  \textbf{Figure~\ref{spectra}} (continued)
\end{figure*}


\section{Spectral energy distributions of our NLSy1 sample}
\label{sec:SED}

\begin{figure*}
  \centering
  \includegraphics[height=6.6cm,trim=15 10 45 25,clip=true]{IRAS13349_best_model.pdf}
  \hspace{3mm}
  \includegraphics[height=6.6cm,trim=15 10 45 25,clip=true]{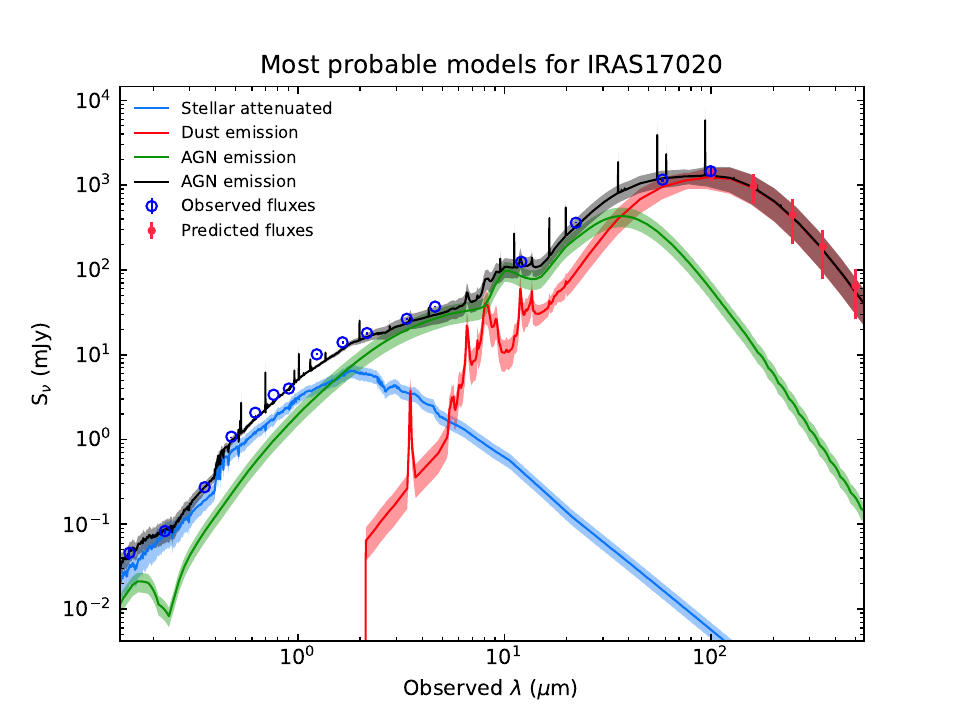} \\
  \vspace{3mm}
  \includegraphics[height=6.6cm,trim=15 10 45 25,clip=true]{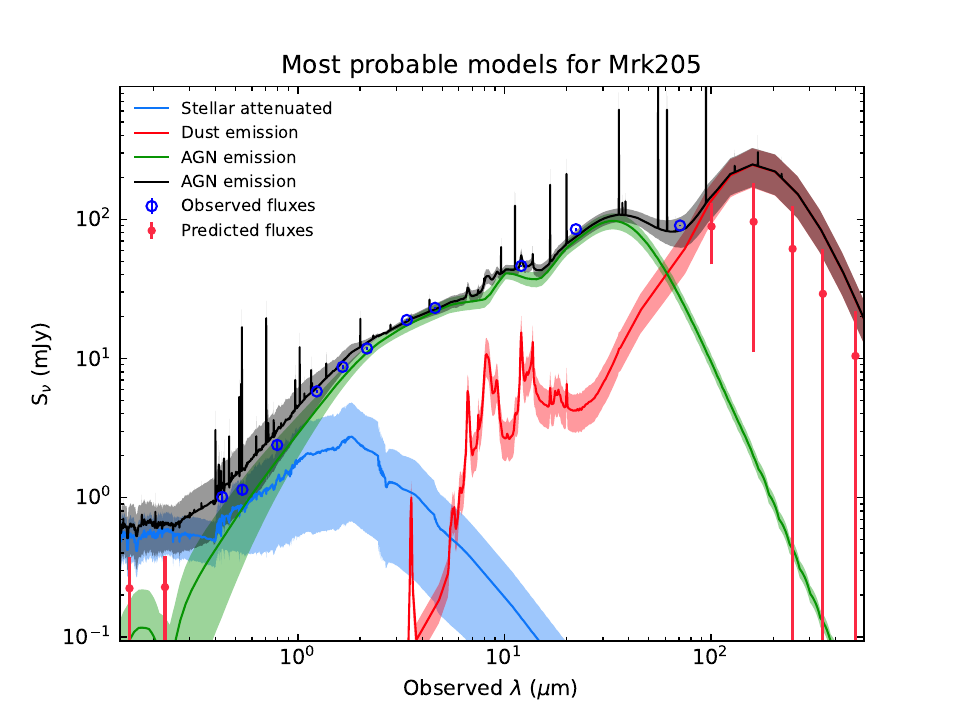}
  \hspace{3mm}
  \includegraphics[height=6.6cm,trim=15 10 45 25,clip=true]{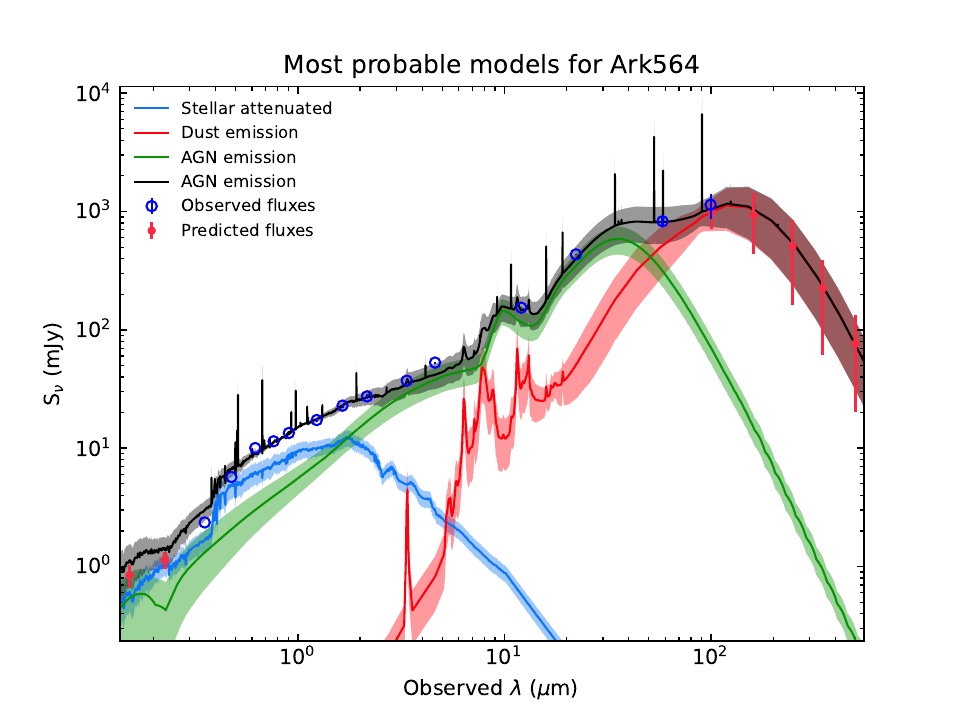} \\
  \vspace{3mm}
  \includegraphics[height=6.6cm,trim=15 10 45 25,clip=true]{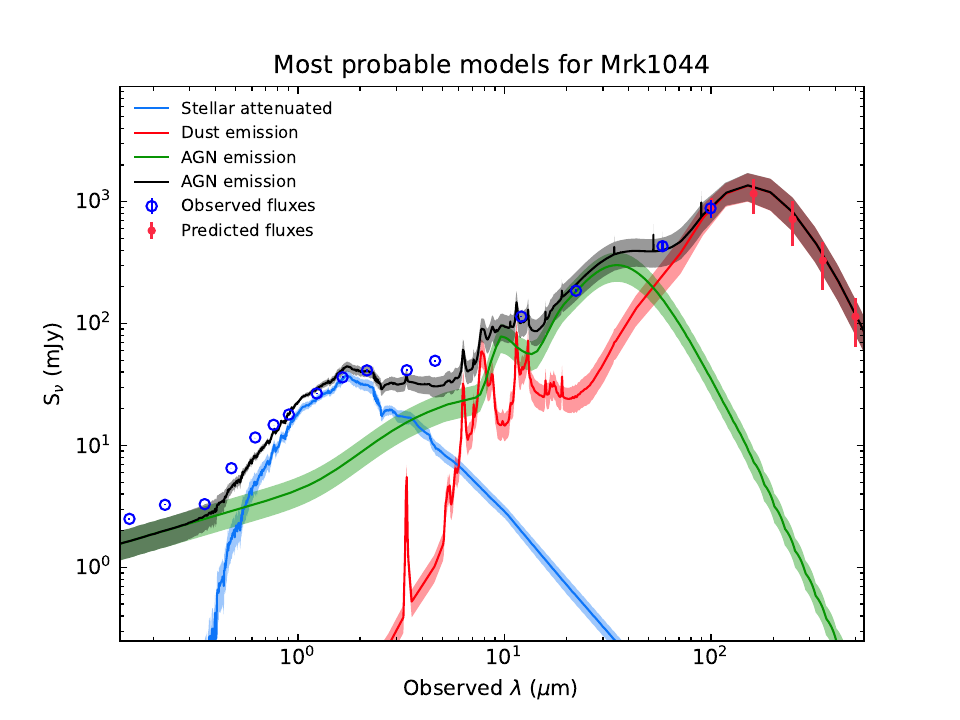}
  \hspace{3mm}
  \includegraphics[height=6.6cm,trim=15 10 45 25,clip=true]{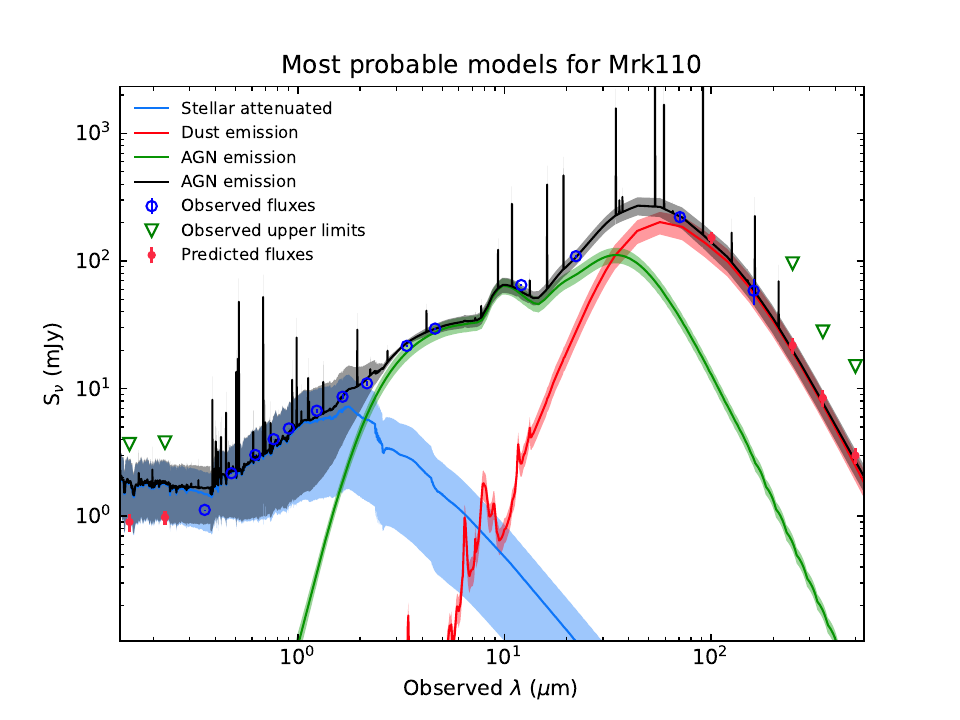}
  \caption{\label{cigale-all} SED for each galaxy in our sample as derived with CIGALE. The green arrows indicate upper limits. The other colored symbols and shaded areas follow Figure \ref{cigale}. Modeled fluxes for the GALEX bands are shown if there is no observations/detection (see Section \ref{sec:cigale}).}
\end{figure*}

\begin{figure*}
  \centering
  \includegraphics[height=6.6cm,trim=15 10 45 25,clip=true]{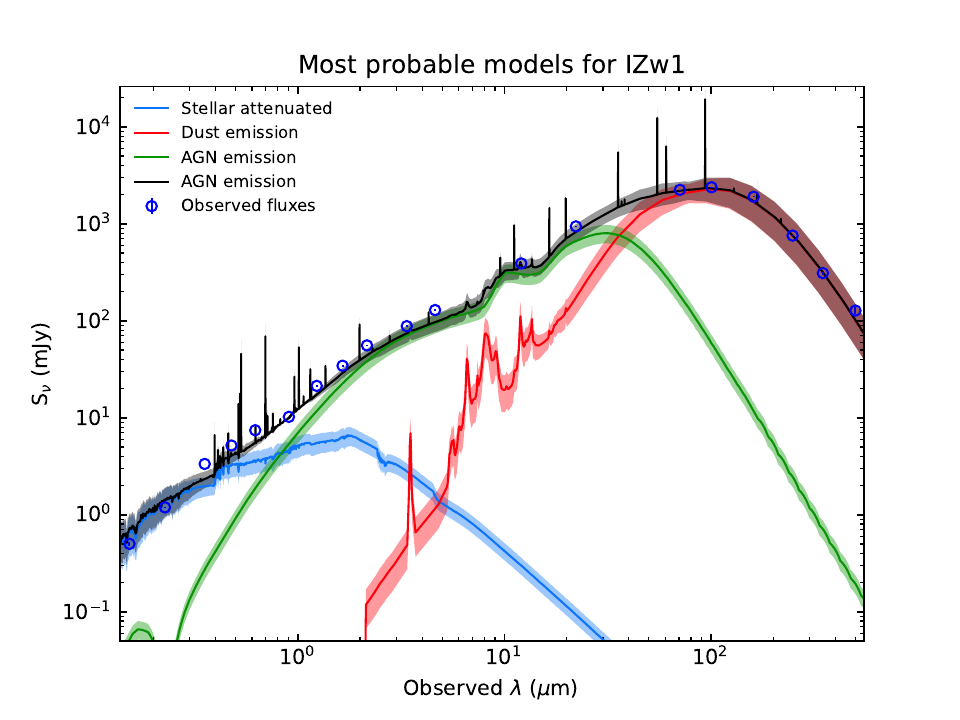}
  \hspace{3mm}
  \includegraphics[height=6.6cm,trim=15 10 45 25,clip=true]{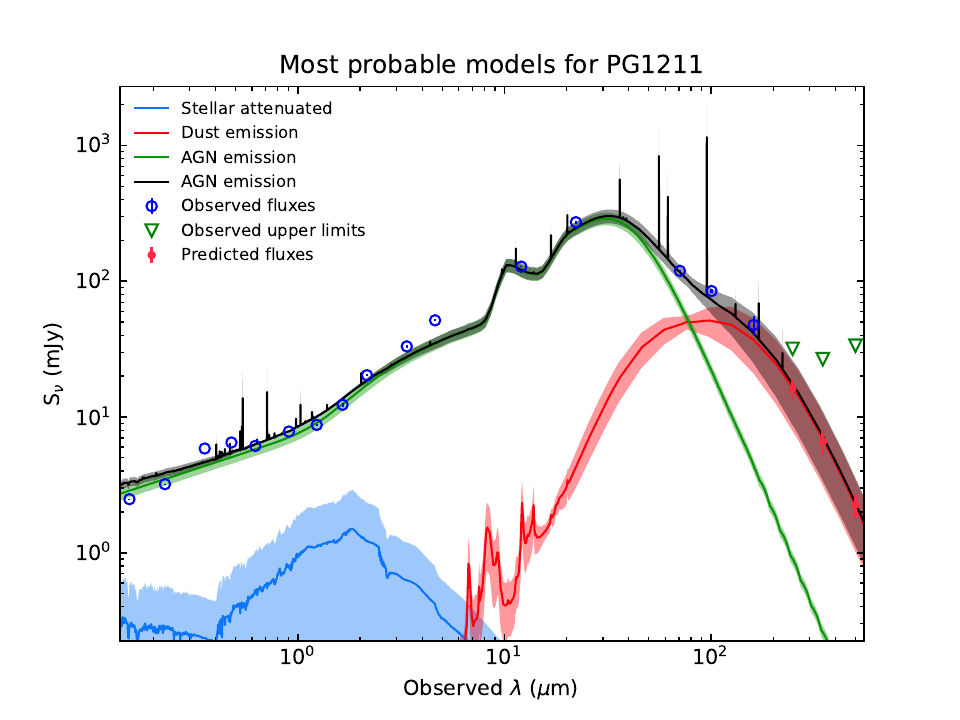} \\
  \vspace{3mm}
  \includegraphics[height=6.6cm,trim=15 10 45 25,clip=true]{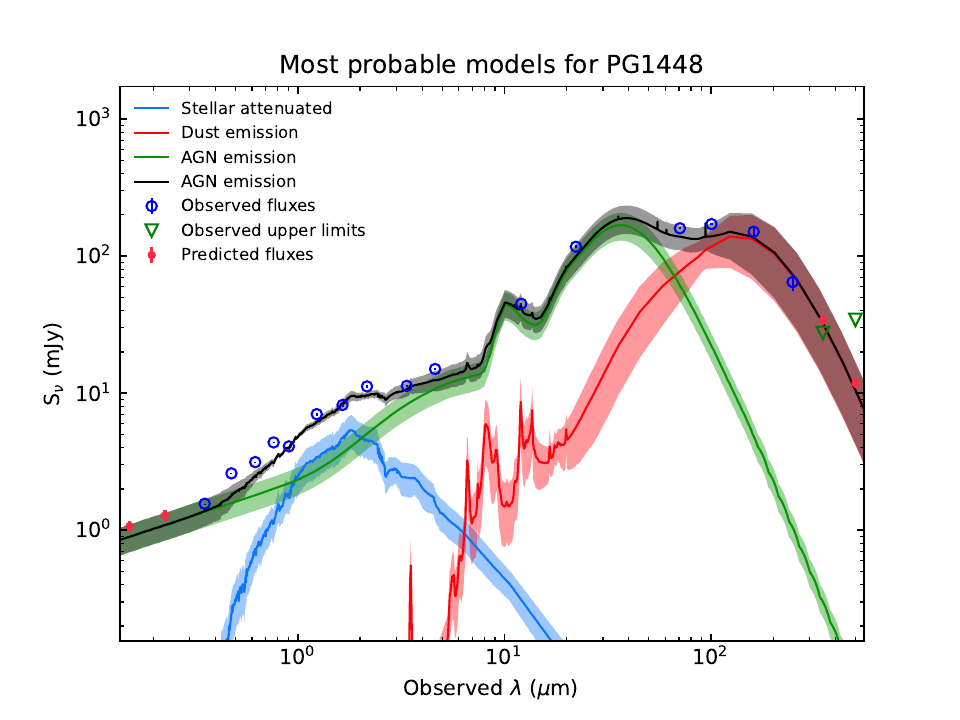}
  \hspace{3mm}
  \includegraphics[height=6.6cm,trim=15 10 45 25,clip=true]{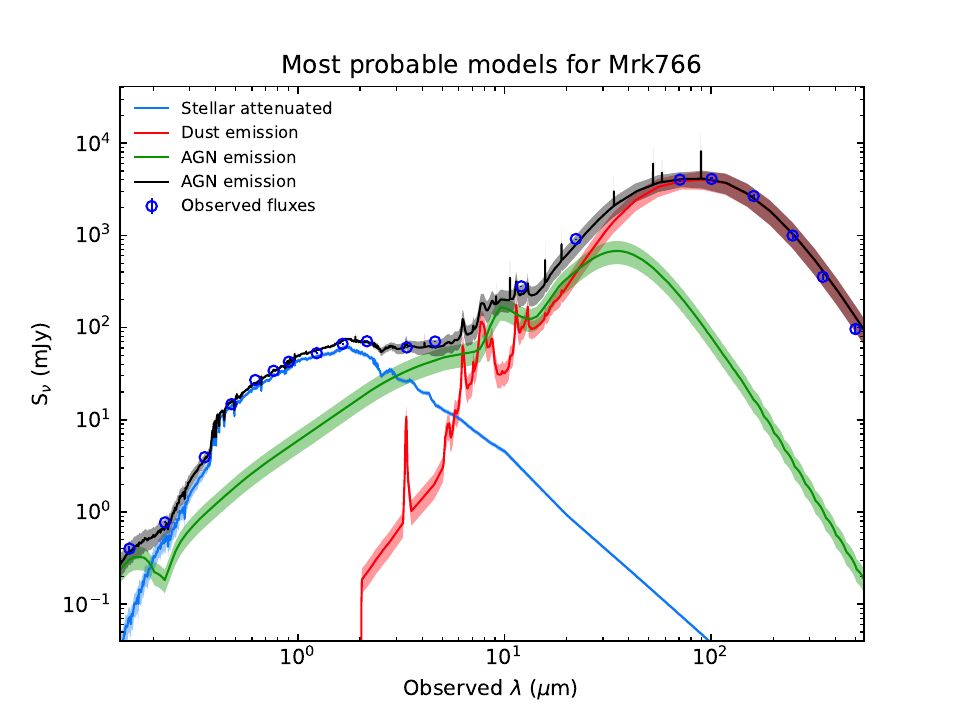} \\
  \textbf{Figure~\ref{cigale-all}} (continued)
\end{figure*}

\label{lastpage}
\end{document}